\documentclass[preprintnumbers,twocolumn]{revtex4}

\usepackage{graphicx}
\usepackage{epsf}
\usepackage{amsmath,amssymb}
\usepackage{color}

\begin{document}
\newcommand{\Hes}{{}^6\textrm{He}}
\newcommand{\Hee}{{}^8\textrm{He}}

\preprint{KUNS-2814, NITEP 68}
\title{Microscopic calculation of proton and alpha-particle inelastic scattering 
to study the excited states of $^{6}$He and $^{8}$He}

\author{Yoshiko Kanada-En'yo}
\affiliation{Department of Physics, Kyoto University, Kyoto 606-8502, Japan}
\author{Kazuyuki Ogata} 
\affiliation{Research Center for Nuclear Physics (RCNP), Osaka University,
  Ibaraki 567-0047, Japan}
\affiliation{Department of Physics, Osaka City University, Osaka 558-8585,
  Japan}
\affiliation{
Nambu Yoichiro Institute of Theoretical and Experimental Physics (NITEP),
   Osaka City University, Osaka 558-8585, Japan}

\begin{abstract}
Elastic and inelastic cross sections of the $p+{}^6\textrm{He}$, $p+{}^8\textrm{He}$, and $\alpha+{}^8\textrm{He}$ reactions 
were investigated using the Melbourne $g$-matrix folding approach
with the theoretical densities of $\Hes$ and ${}^8\textrm{He}$ obtained by a microscopic structure model of 
antisymmetrized molecular dynamics (AMD). 
Microscopic coupled-channel (MCC) calculations of the $p+{}^6\textrm{He}$ and $p+{}^8\textrm{He}$ reactions
were performed to investigate transition properties of the ${}^6\textrm{He}(2^+_1)$ and ${}^8\textrm{He}(2^+_1)$ states.
The MCC+AMD calculations reproduced elastic cross sections of the $p+{}^6\textrm{He}$ reaction
at $E=40.9$~MeV/A  and of the $p+{}^8\textrm{He}$ reaction at $E=32.5$ and 72~MeV/A, which have both been 
measured by inverse kinematics experiments.
For $p+{}^6\textrm{He}$ inelastic scattering to the $2^+_1$ state,   
the calculated result was in reasonable agreement with the $(p,p')$ data 
at $E=24.5$ and 40.9~MeV/A and supported the AMD prediction of the neutron transition matrix element
$M_n=7.9$ fm$^2$.
For the $p+{}^8\textrm{He}$ inelastic scattering to ${}^8\textrm{He}(2^+_1)$, the MCC+AMD calculation
overshot the $(p,p')$ cross sections at $E=72$~MeV/A by a factor of three.
According to a phenomenological model analysis, 
$M_n$ values in the range of 4--6~fm$^2$ were
favored to reproduce the ${}^8\textrm{He}(2^+_1)$ cross sections
of the $p+{}^8\textrm{He}$ reaction at $E=72$~MeV/A.
For the $\alpha+{}^8\textrm{He}$ reaction,  
the MCC+AMD calculation reproduced the elastic cross sections 
at $E=26$~MeV/A. Theoretical predictions of the ($p,p'$) and ($\alpha,\alpha'$) cross sections to 
the ${}^8\textrm{He}(0^+_2)$, ${}^8\textrm{He}(1^-_1)$, ${}^8\textrm{He}(2^+_3)$ and ${}^8\textrm{He}(3^-_1)$ states are also given.
\end{abstract}

\maketitle

\section{Introduction}
Elastic and inelastic $(p)$ cross sections have been extensively measured 
for various stable nuclei to investigate the density of the ground states 
and the transition profiles of nuclear excitations. 
Because of their higher sensitivity of the $p$ scattering to the neutron part, 
the $(p,p')$ data for $Z\ne N$ stable nuclei have often been used 
to determine the neutron transition matrix elements $M_n$. 
as the counterparts to electric probes for the proton transition matrix elements $M_p$.
Additionally, the $(\alpha,\alpha')$ reaction is an alternative tool 
that can be used to observe the isoscalar component of low-lying nuclear excitations
as well as isoscalar giant resonances (e.g., see 
Refs.~\cite{VanDerBorg:1981qiu,Youngblood:1999zza,Harakeh-textbook,Uchida:2004bs} and the refs.
therein). 
Because of the existence of high-quality beams, 
precise data on the $p$ and $\alpha$ scattering are available 
for various stable nuclei in wide ranges of incident energies. 
To investigate the 
isoscalar and isovector components of $2^+$ excitations, 
the neutron-proton ratio $M_n/M_p$ of $E2$ transitions
has been previously evaluated using a combination of $\gamma$-decay and the inelastic scattering data on
different probes including $e$, $p$, $\alpha$, and $\pi^+/\pi^-$,
for a wide range of mass numbers~\cite{Bernstein:1977wtr,Bernstein:1979zza,Bernstein:1981fp,Brown:1980zzd,Brown:1982zz}.
Over the years, $\alpha$ inelastic scattering experiments
have also been used  as a probe to cluster states 
in $Z=N$ nuclei, such as those of $^{12}$C and $^{16}$O \cite{Wakasa:2006nt,Itoh:2011zz,Kawabata:2013xea,Adachi:2018pql}. 

For unstable nuclei, pion and electron scattering experiments are not practical.
Instead, the $p$ scattering off unstable nuclei has generally been investigated by experiments using 
inverse kinematics. 
The amount of $p$ elastic and inelastic scattering data, which provides 
useful information on the structure of unstable nuclei, 
is rapidly increasing.
Further, $\alpha$ scattering experiments for unstable nuclei
have also been achieved  using inverse kinematics~\cite{Wolski:2002gzz,Wolski:2003moy,Furuno:2019lyp}.

On the theoretical side, 
describing the existing $(p,p')$ and $(\alpha,\alpha')$ data
with reaction calculations is an urgent issue for the study of excited states of target nuclei. 
For analyses of inelastic scattering, 
phenomenological reaction models, such as distorted wave
Born approximation (DWBA),
are often performed by adjusting nucleon$-$nucleus and nucleus$-$nucleus potentials to elastic scattering data.
Such phenomenological tuning, which usually depends on the incident energy and the target nuclei, 
requires a significant amount of 
elastic scattering data. However, for unstable nuclei, data are available only for limited energy and angle ranges and are not as high-quality as  data on stable nuclei. 
Given this limit, the development of 
a microscopic reaction model, that does not rely on system- and energy-dependent tuning is necessary.

For experimental and theoretical studies of 
$p$ scattering off unstable nuclei, microscopic reaction approaches using the $g$-matrix folding model
have been developed.
These approaches are suitable for our purpose
because the energy dependence and medium effects are taken into account 
in the effective $g$-matrix nucleon-nucleon~($NN$) interaction, which is derived from 
a realistic $NN$ force based on the Brueckner theory.
For studies on the $p$ inelastic scattering,  
the Jeukenne-Lejeune-Mahaux (JLM) interaction \cite{Jeukenne:1977zz} has often been used 
as an effective $g$-matrix $NN$ interaction and has been applied to microscopic calculations of $(p,p')$ reactions 
for unstable nuclei including He, Li, Be, and C isotopes, along with
microscopic structure calculations of the target nuclei~
\cite{Lapoux:2001kpc,Skaza:2005uff,Jouanne:2005pb,Takashina:2005bs,Takashina:2008zza,Lapoux:2015jva,Matsumoto:2017mau,Ogawa:2020qtt}.

Since the 1990s, experimental studies of $p+\Hes$ and $p+\Hee$ reactions have been widely 
performed~(see the review in Ref.~\cite{Lapoux:2015jva}). 
To extract information about the ground and  $2^+_1$  states
via $(p,p)$ and $(p,p')$ reactions, respectively, 
a systematic analysis using JLM has been performed 
 using the diagonal and transition densities of $\Hes$ and $\Hee$ obtained using
theoretical structure models.
$p+\Hes$ inelastic scattering has also been investigated 
using a four-body calculation with a continuum discretized coupled channel
considering the higher-order effects of continuum coupling with 
all $0^+$, $1^-$, and $2^+$ partial waves \cite{Ogawa:2020qtt}.
In principle, the $g$-matrix folding approach should not contain adjustable parameters, but 
JLM reaction calculations require phenomenological parameter tuning 
to fit the elastic data, which is usually performed by introducing renormalization factors of potentials. 

Along a similar line to the microscopic reaction approaches, the Melbourne group has 
developed a microscopic folding model with 
an improved $g$-matrix $NN$ interaction and demonstrated its success in reproducing 
$p$+nucleus elastic scattering for a wide range of target mass numbers and incident energies 
\cite{Amos:2000}. 
The Melbourne $g$-matrix folding approach has been widely tested
for $p$ and $\alpha$ elastic scattering, and the framework has also been successfully applied to 
inelastic 
processes~\cite{Amos:2000,Lagoyannis:2000te,Stepantsov:2002efb,Karataglidis:2007yj,Minomo:2009ds,Toyokawa:2013uua,Minomo:2017hjl,Egashira:2014zda,Minomo:2016hgc,Kanada-Enyo:2019prr,Kanada-Enyo:2019qbp,Kanada-Enyo:2019uvg,Kanada-Enyo:2020zpl}.
Since the approach with the Melbourne $g$-matrix $NN$ interaction has no adjustable parameters, 
it is a useful tool to test the reliability of structure inputs without model ambiguity on the reaction side. 
Because of its merit, 
one can directly access the structural properties of the ground and excited states via $p$ scattering. 

In this study, we investigated the elastic and inelastic cross sections of the $p+\Hes$, $p+\Hee$, and $\alpha+\Hee$ reactions with the Melbourne $g$-matrix folding approach using the 
densities of $\Hes$ and $\Hee$ obtained by a structure model of 
antisymmetrized molecular dynamics (AMD) \cite{KanadaEnyo:1995tb,KanadaEnyo:1995ir,KanadaEn'yo:2012bj}.
In the present calculations, 
we adopted the same framework of the reaction approach as 
in Refs.~\cite{Kanada-Enyo:2019prr,Kanada-Enyo:2019qbp,Kanada-Enyo:2019uvg,Kanada-Enyo:2020zpl}.
Namely, we used the folding model 
with a simplified treatment of the exchange term of the optical potential
and multiple scattering theory \cite{Ker59} for $p$ scattering and 
applied the approximation of an extended version of 
the nucleon$-$nucleus folding model for $\alpha$ scattering. 

For the structural calculation using the AMD framework,
the proton and neutron components of the diagonal and transition densities of 
$\Hes$ and $\Hee$ were microscopically obtained.
Using the AMD densities, 
the microscopic coupled-channel (MCC) calculations of the $p+\Hes$ and $p+\Hee$ reactions 
were performed.
Using this MCC+AMD model, the $p+\Hes$, $p+\Hee$, and $\alpha+\Hee$ reactions can be investigated
on the same footing. We first checked the ability of the MCC+AMD calculations to reproduce 
 elastic cross sections by comparison with the $(p,p)$ data and then investigated the transition properties of the $\Hes(2^+_1)$ and $\Hee(2^+_1)$ states 
via the $(p,p')$ cross sections.
The $\alpha+\Hee$ elastic and inelastic cross sections were also calculated using the MCC+AMD model,
and the results were compared with the $(\alpha,\alpha)$ data observed at $E=72$~MeV/A.
Further, theoretical predictions of the inelastic cross sections to 
the $\Hee(0^+_2)$, $\Hee(1^-_1)$, and $\Hee(3^-_1)$ states were made.

The rest of the study is organized as follows. In Sec.~\ref{sec:reaction},
the reaction model using the Melbourne $g$-matrix folding approach is explained.
In Sec.~\ref{sec:AMD}, the AMD calculation of the structures of the He isotopes is described. 
Sec.~\ref{sec:he6p} presents the results of the $p+\Hes$ reaction, and Sec.~\ref{sec:he8p} discusses 
the  $p+\Hee$ scattering to the $\Hee(0^+_1)$ and $\Hee(2^+_1)$ states.
The calculated results of the $p+\Hee$ and $\alpha+\Hee$ scattering 
to higher excited states are shown in Sec.~\ref{sec:he8-J0123}. 
Finally, a summary is given in Sec.~\ref{sec:summary}.

\section{Reaction model: Melbourne $g$-matrix folding approach} \label{sec:reaction}

The reaction calculations of $p$ and $\alpha$ scattering were performed
using the Melbourne $g$-matrix folding model approach, 
as in  
Refs.~\cite{Kanada-Enyo:2019prr,Kanada-Enyo:2019qbp,Kanada-Enyo:2019uvg,Kanada-Enyo:2020zpl}. 

The nucleon$-$nucleus potentials were calculated 
by folding the Melbourne $g$-matrix $NN$ interaction with the diagonal and transition densities of target nuclei, 
which were microscopically obtained by AMD.
The Melbourne $g$-matrix interaction is an effective $NN$ interaction in a nuclear medium
and is based on a bare $NN$ interaction of the Bonn B potential~\cite{Machleidt:1987hj}.
The $\alpha$$-$nucleus potentials were obtained by folding the 
nucleon$-$nucleus potentials with an $\alpha$ density of a Gaussian form. 
In the present calculation,
 the spin-orbit term of the $p$$-$nucleus potentials was not taken into account 
to avoid complexity. It should be noted that
the spin-orbit interaction can smear the
dip structure of elastic cross sections,
but it gives a minor contribution to absolute peak amplitudes at incident energies lower 
than $E_p=$100 MeV except for at backward angles.
For the details of the present reaction model, see
Refs.~\cite{Kanada-Enyo:2019prr,Kanada-Enyo:2019qbp,Kanada-Enyo:2019uvg,Kanada-Enyo:2020zpl}
 and the refs. therein.

\section{Structure calculations and properties of $\Hes$ and $\Hee$} \label{sec:AMD}

\subsection{AMD calculations}
For use in MCC calculations, the diagonal and transition densities of the target nuclei 
were calculated using
the AMD wave functions of $\Hes$ and $\Hee$, which were
obtained in a previous study of He isotopes~\cite{Kanada-Enyo:2007iri}. 
We used the wave functions labeled as 
``m56'' and ``v58'', 
which represented two choices of parametrization given in Table 1 of Ref.~\cite{Kanada-Enyo:2007iri}. 
In this study, we call the former AMDm56 and the latter AMDv58.
The details of the calculation procedures and the ground state structures 
of $\Hes$ and $\Hee$ were described in the previous work. 
It should be stressed that 
the AMD is a microscopic structure model that 
considers the degrees of freedom of all the nucleons. Even though the model does not assume any clusters, 
$\alpha$-like cluster structures with valence neutrons were obtained by the AMD calculation 
for low-lying $\Hes$ and $\Hee$ states.

In the present structure calculation, we used the basis AMD configurations obtained in the previous
calculation but improved the accuracy of 
the numerical integration of the angular momentum projection
to calculate the diagonal and transition densities of the excited states with high accuracy precision. 
Using the updated wave functions, 
we perform a diagonalization of the Hamiltonian and norm matrices and 
recalculated the structural properties including the energies, radii, and transition strengths.

\subsection{Properties of the $0^+_1$ and $2^+_1$ states of $\Hes$ and $\Hee$}  \label{sec:AMD-results}

\begin{table*}[!ht]
\caption{Rms point-proton ($R_p$) and point-neutron ($R_n$) radii, 
$2^+_1$ excitation energies $(E_x)$, and proton $(M_p)$ and neutron $(M_n)$ components of the 
$0^+_1\to 2^+_1$ transition matrix elements of 
$\Hes$ and $\Hee$. The results of AMDm56 and AMDv58 
are shown together with 
the experimental data~\cite{Wang:2004ze,Mueller:2008bj,Tanihata:1992wf,Ozawa:2001hb,Tilley:2002vg,Tilley:2004zz}. 
Theoretical values of other models, including those of the stochastic variational calculation~\cite{Varga:1994fu}, the NCSM
calculation of Ref.~\cite{Caurier:2005rb} and that of Refs.~\cite{Navratil:2002zz,Navratil:2003ef}, are shown for comparison. 
In the rightmost column, 
the theoretical values of the halo-type density set of
$\rho_\textrm{halo}$~(NCSM-based halo-type diagonal density) and 
$\rho^\textrm{tr}_\textrm{Tassie}$~(2pF-Tassie transition density) for $\Hes$ from Ref.~\cite{Lapoux:2015jva} are shown. 
The experimental values of $R_p$ were extracted from the charge radii~\cite{Wang:2004ze,Mueller:2008bj},
and those of $R_n$ were obtained from the $R_p$ data and the empirical matter radii in Refs.~\cite{Tanihata:1992wf,Ozawa:2001hb}. 
For NCSM\cite{Navratil:2002zz,Navratil:2003ef}, 
the values of the NCSM(10$\hbar\omega$) calculation of $\Hes$ and those of  the NCSM$_{v3\textrm{eff}}$(4$\hbar\omega$) calculation of $\Hee$
were taken from Ref.~\cite{Lapoux:2015jva}. 
 \label{tab:radii}
}
\begin{center}
\begin{tabular}{lcccccccc}
\hline
 	&	Expt.	&	AMDm56	&	AMDv58	&	SVM	&	NCSM\cite{Caurier:2005rb} 	&	NCSM\cite{Navratil:2003ef,Navratil:2002zz} & 
halo-type\cite{Lapoux:2015jva}	\\
$\Hes$	&		&		&		&		&		&		&		\\
$R_p$ (fm)	&	1.92(1)	&	1.92	&	1.79	&	1.80	&	1.76(3)	&	1.76	&	2.03 	\\
$R_n$ (fm)	&	2.39--2.77	&	2.48	&	2.39	&	2.67	&	2.55(10)	&	2.36	&	2.72 	\\
$E_x(2^+_1)$ (MeV)	&	1.80	&	1.4 	&	2.0 	&		&		&	2.63	&		\\
$M_p$ (fm$^2$)	& 		&	0.90 	&	0.92 	&		&		&	1.03	&	2.44 	\\
$M_n$ (fm$^2$)	&		&	7.9 	&	7.9 	&		&		&	7.73	&	7.80 	\\
	&		&		&		&		&		&		&		\\
$\Hee$	&		&		&		&		&		&		&		\\
$R_p$ (fm)	&	1.81(3)	&	1.92	&	1.69	&	1.71	&	1.74(6)	&	2.00	&		\\
$R_n$ (fm)	&	2.60--2.75	&	2.63	&	2.33	&	2.53	&	2.60(10)	&2.59		&		\\
$E_x(2^+_1)$ (MeV)	&	3.1(5)	&	4.1 	&	6.1 	&		&		&		&		\\
$M_p$ (fm$^2$)	&		&	0.32 	&	0.38 	&		&		&	0.50 	&		\\
$M_n$ (fm$^2$)	&		&	7.6 	&	6.4 	&		&		&	6.67 	&		\\
\hline
\end{tabular}
\end{center}
\end{table*}

\begin{figure}[!htpb]
\includegraphics[width=9 cm]{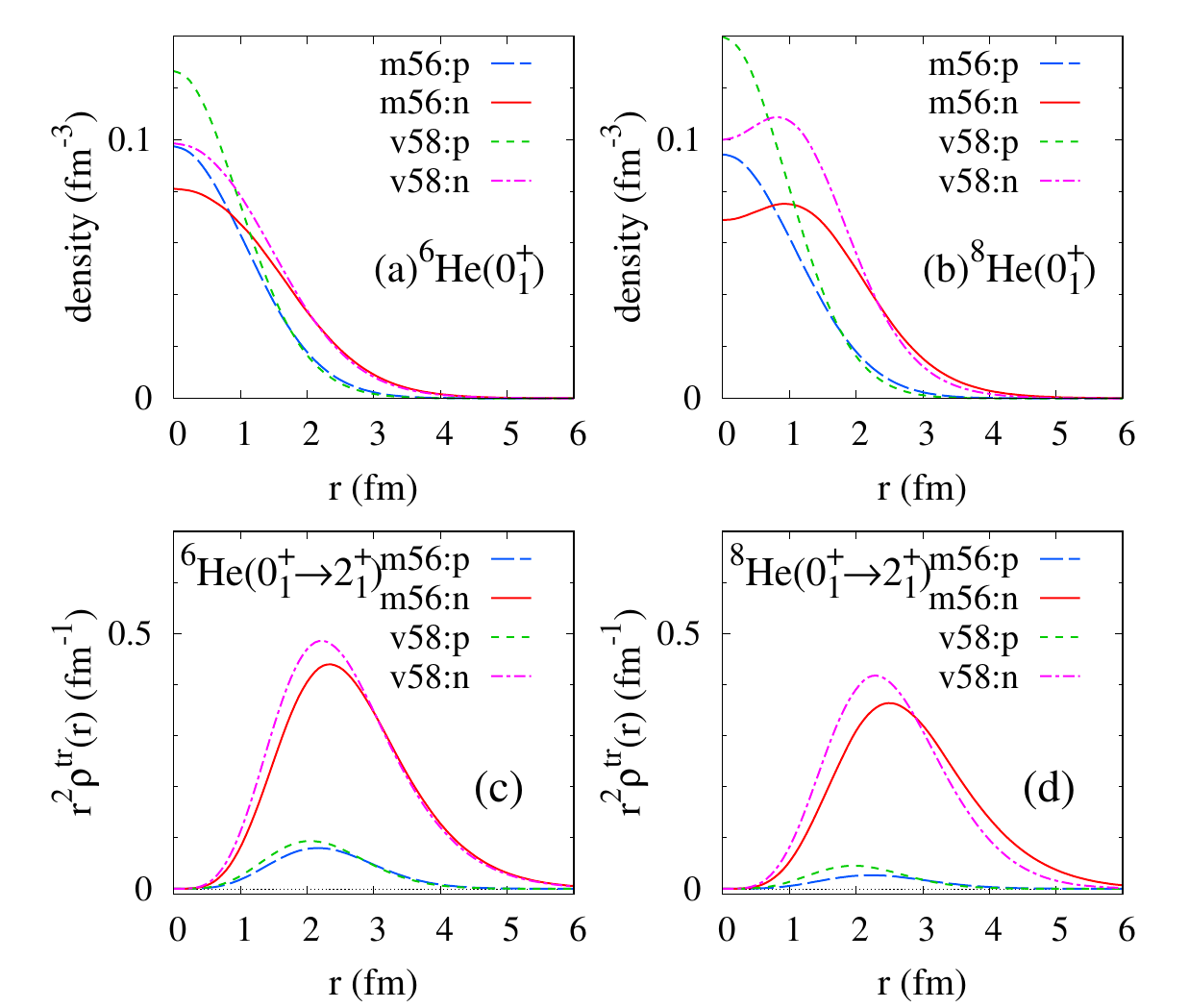}
  \caption{Diagonal and transition densities calculated using AMDm56 and AMDv58. 
Proton and neutron components of 
the diagonal densities for  (a) $\Hes (0^+_1)$ and  (b) $\Hee (0^+_1)$
and those of the $0^+_1\to 2^+_1$ transition densities for  (c) $\Hes$ and (d) $\Hee$.
The $r^2$-weighted transition densities are plotted.
	\label{fig:trans-he}}
\end{figure}

The structural properties of $\Hes$ and $\Hee$ are summarized in 
Table \ref{tab:radii}. 
The calculated values of the root-mean-square~(rms) point-proton and point-neutron
radii of the ground state, $2^+_1$ energies, and the proton  
and neutron  components of the
$0^+_1\to 2^+_1$ transition matrix elements of $\Hes$ and $\Hee$ are shown together with the experimental radii and
energies. 
The proton and neutron components of the diagonal and transition densities are shown in Fig.~\ref{fig:trans-he}.
In both the AMDm56 and AMDv58 results, the proton density remained in the inner region, whereas 
the neutron density was distributed in the outer region indicating 
a neutron halo structure in  $\Hes(0^+_1)$ and a neutron skin structure in $\Hee(0^+_1)$, which are generated by
loosely bound valence neutrons around the $\alpha$ core.
The AMDm56 and AMDv58 results were qualitatively similar
but the AMDv58 results showed smaller radii than the AMDm56 results
for $\Hee$ in particular. As shown in Fig.~\ref{fig:trans-he}(b), 
the AMDv58 results also showed a weaker neutron skin than did the AMDm56 results (Fig.~\ref{fig:trans-he}(b)). 
This interaction dependence is regarded as an
ambiguity of the structure calculation using the AMD model.

For the $0^+_1\to 2^+_1$ transition, a remarkable neutron dominance was obtained in both the 
$\Hes$ and $\Hee$ systems. 
This neutron dominance in the $\Hes(2^+_1)$ and $\Hee(2^+_1)$ excitations 
is a general phenomenon in proton-closed nuclei with $N>Z$ and is caused by valence neutrons
around a core, as seen in $^{18}\textrm{O}$.
The neutron transition densities demonstrated a single-peak structure at the nuclear surface,
as shown in Figs.~\ref{fig:trans-he}(c) and (d).

To reveal the structure model ambiguity with respect to the proton and neutron densities, 
we compared the AMD results with several other calculations. 
In Table \ref{tab:radii}, we show the theoretical values obtained by 
the stochastic variational calculation~\cite{Varga:1994fu} and the no-core shell model~(NCSM) calculations 
from Refs.~\cite{Caurier:2005rb,Navratil:2003ef,Navratil:2002zz}. For $\Hes$, we also show 
the radii and transition matrix elements for a halo-type density set of 
the NCSM-based halo-type diagonal density~($\rho_\textrm{halo}$)  
and the 2pF-Tassie transition density~($\rho^\textrm{tr}_\textrm{Tassie}$) from
Ref.~\cite{Lapoux:2015jva}.
All of the models gave qualitatively similar results with respect to the halo and skin structures, but quantitative differences can be observed in the neutron radii particularly. 
Regarding the $0^+_1\to 2^+_1$ transition properties,  the NCSM calculation gave almost the same values of $M_p$ and $M_n$ 
as the AMD predictions for both the $\Hes$ and $\Hee$ systems.

\subsection{Higher excited states of $\Hee$}

\begin{figure}[!htpb]
\includegraphics[width=8 cm]{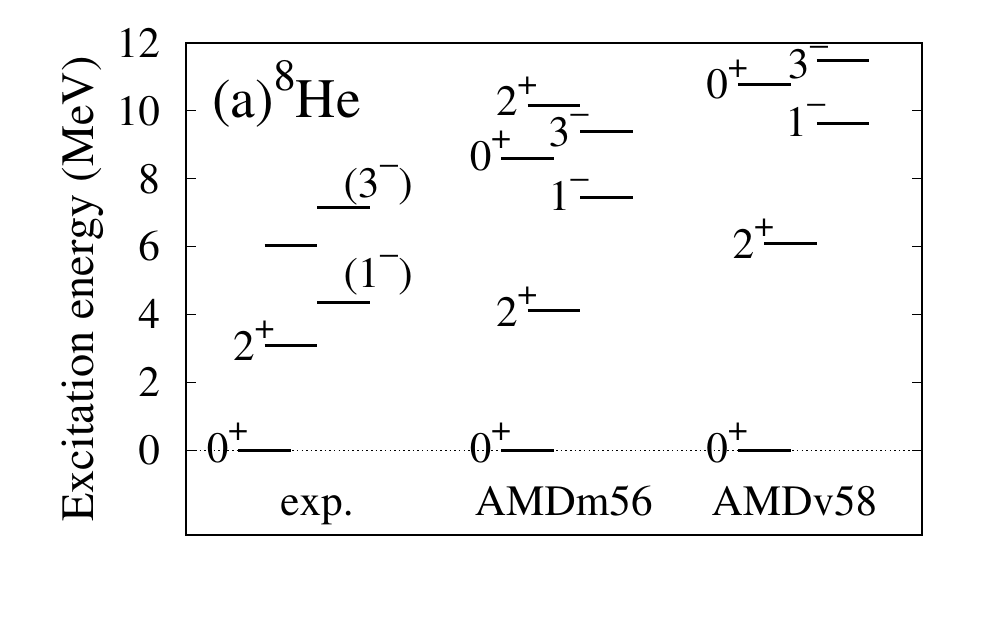}
  \caption{Calculated energy spectra obtained by AMDm56 and AMDv58 and the experimental data from Ref.~\cite{Tilley:2004zz}.
	\label{fig:spe-he8}}
\end{figure}

\begin{table}[!ht]
\caption{Excitation energies, rms radii, and $M_p$ and $M_n$ values of the
$0^+_1\to J^\pi$ transitions of $\Hee$ calculated using AMDm56. The units for energy and radius are
MeV and fm, respectively. The units for 
$M_p$ and $M_n$  for the IS0 and IS1 transitions are fm$^{\lambda+2}$, and those  
for the $E2$ and $E3$ transitions are fm$^{\lambda}$.
 \label{tab:he8-J0123}
}
\begin{center}
\begin{tabular}{ccccccccc}
\hline
$\Hee(J^\pi)$&$E_x$	&	$R_p$&	$R_n$	&	$R_m$	&		&	$M_p$	&	$M_n$	\\
$0^+_1$	&	0.0 	&	1.92 	&	2.63 	&	2.47 	&		&		&		\\
$0^+_2$	&	8.6 	&	2.12 	&	3.10 	&	2.89 	&	IS0	&	0.51 	&	2.8 	\\
$2^+_1$	&	4.1 	&	1.98 	&	2.78 	&	2.60 	&	$E2$	&	0.32 	&	7.6 	\\
$2^+_2$	&	10.2 	&	2.05 	&	2.99 	&	2.78 	&	$E2$	&	$-0.01$	&	0.04 	\\
$2^+_3$	&	13.4 	&	2.43 	&	3.70 	&	3.43 	&	$E2$	&	0.30 	&	4.5 	\\
$1^-_1$	&	7.4 	&	2.08 	&	3.30 	&	3.04 	&	IS1	&	$-1.84$	&	13.5 	\\
$3^-_1$	&	9.4 	&	2.08 	&	3.24 	&	2.99 	&	$E3$	&	$-0.04$	&	37 	\\
\hline
\end{tabular}
\end{center}
\end{table}

In  Fig.~\ref{fig:trans-he}, the calculated energy spectra of $\Hee$, including the excited states above the $2^+_1$ state
are shown together with the experimental spectra.
The calculated values of the proton, neutron, and matter 
radii of the ground and excited states, and the proton and neutron components of the
transition matrix elements from the ground to excited states are listed in Table~\ref{tab:he8-J0123}.
Similar to the $\Hee(2^+_1)$ state, the $\Hee(0^+_2)$, $\Hee(1^-_1)$, $\Hee(2^+_3)$, and $\Hee(3^-_1)$ states also had neutron skin structures and 
neutron-dominant transitions, which again indicate the 
predominant contribution of valence neutrons around the $\alpha$-like cluster. 
More details on the transition properties for these higher states are discussed in Sec.~\ref{sec:he8-J0123}.


\section{$p+\Hes$ scattering} \label{sec:he6p}
\subsection{MCC+AMD results for $p+\Hes$ cross sections}

\begin{figure}[!htpb]
\includegraphics[width=7 cm]{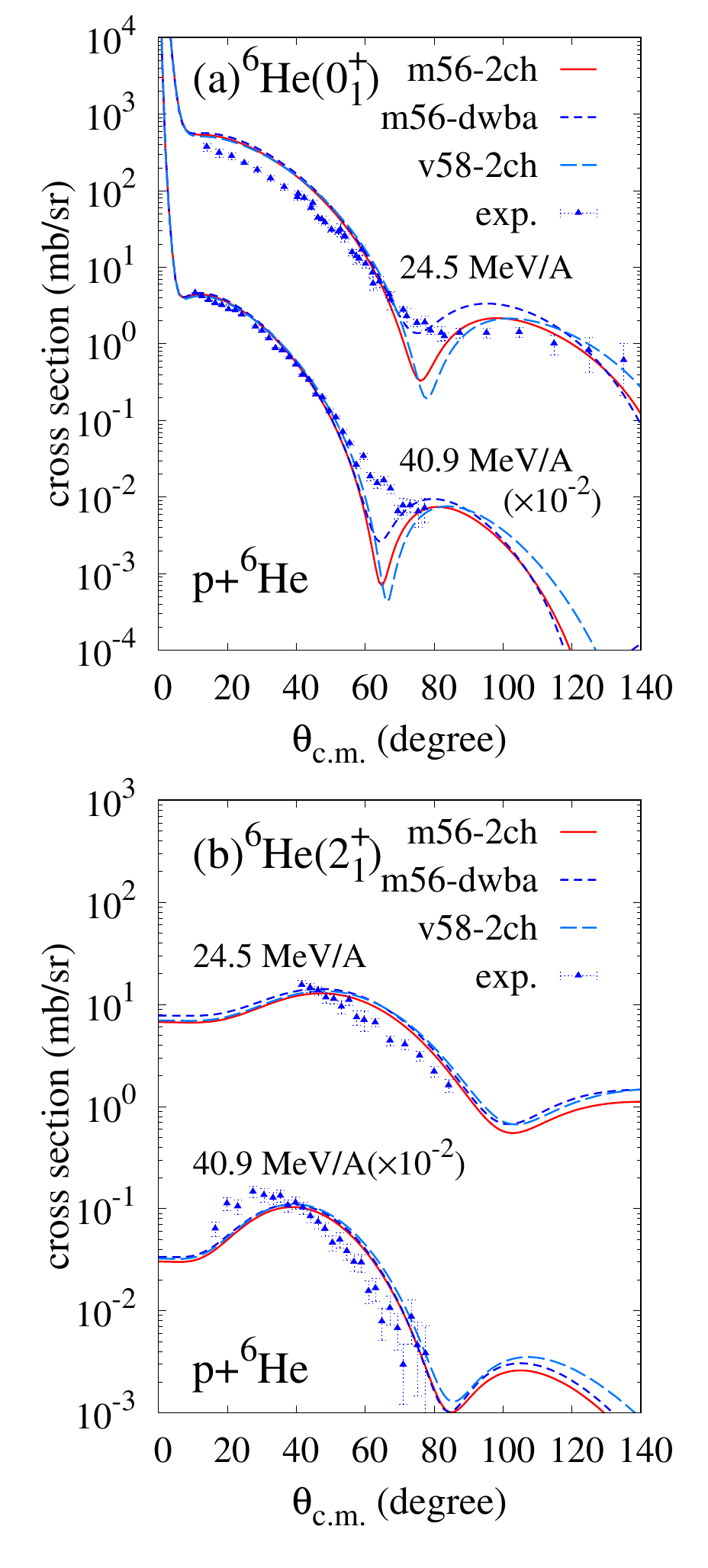}
\caption{Calculated $\Hes(0^+_1)$ and $\Hes(2^+_1)$
cross sections of the $p+\Hes$ scattering 
at $E=24.5$~MeV/A and $E=40.9$~MeV$~(\times 10^{-2})$
compared with the experimental data. 
The 2ch(MCC) calculations using AMDm56 and AMDv58  and 
the DWBA calculation using  AMDm56 are shown. 
The experimental data at $E=24.5$, 25, and 25.7~MeV/A~\cite{Stepantsov:2002efb,Giot:2005zz,Wolski:1999ppw} and 
the data 
at $E=40.9$~MeV/A~$(\times 10^{-2})$~\cite{Lagoyannis:2000te} 
are shown. 
	\label{fig:cross-he6p}}
\end{figure}

The $p+\Hes$ reactions at $E=24.5$ and 40.9~MeV/A 
were calculated using the MCC calculation with the AMDm56 and AMDv58 densities.
The $\Hes(0^+_1)$ and $\Hes(2^+_1)$ states and all of the 
$\lambda=0$ and $\lambda=2$ transitions between them were taken into account.
In addition to this 2ch(MCC) calculation, one-step cross sections were also calculated 
using the DWBA.
In Fig.~\ref{fig:cross-he6p}, 
the calculated $p+\Hes$ cross sections are compared with 
experimental data observed in inverse kinematics~\cite{Stepantsov:2002efb,Lagoyannis:2000te}.

Fig.~\ref{fig:cross-he6p}(a) shows the $p+\Hes$ elastic cross sections. 
The MCC calculations using the AMDm56 and AMDv58 densities 
reasonably described 
the observed $(p,p)$ data at $E=24.5$~MeV/A and did a good job of reproducing the data 
at $E=40.9$~MeV/A
except for the dip structure, 
which could be improved by the smearing effect of the 
spin-orbit interaction omitted in the present calculation.

Unlike the AMDm56 results, 
the second peak of the cross section was shifted to backward angles in the AMDv58 results
because of the smaller radii of $\Hes(0^+_1)$, 
but there was not enough experimental data to select a better calculation. 
Compared with the DWBA~(one-step) cross sections, the MCC calculation obtained 
a lower-amplitude 
second peak because of the coupled channel~(CC) effect with the $2^+_1$ state 
and showed better agreement with the $(p,p)$ data at $E=24.5$~MeV/A. 

Fig.~\ref{fig:cross-he6p}(b) shows the inelastic cross sections to the $\Hes(2^+_1)$ state. 
In both the AMDm56 and AMDv58 results, 
the first-peak amplitude of the cross sections was in reasonable
agreement with the data at $E=24.5$ and 40.9~MeV/A, 
but the calculation did not satisfactorily describe  the behavior of the angle dependence in detail. 
This result indicates that
the neutron transition strength of the AMD prediction is a reasonable value.

\begin{figure}[!htpb]
\includegraphics[width=9 cm]{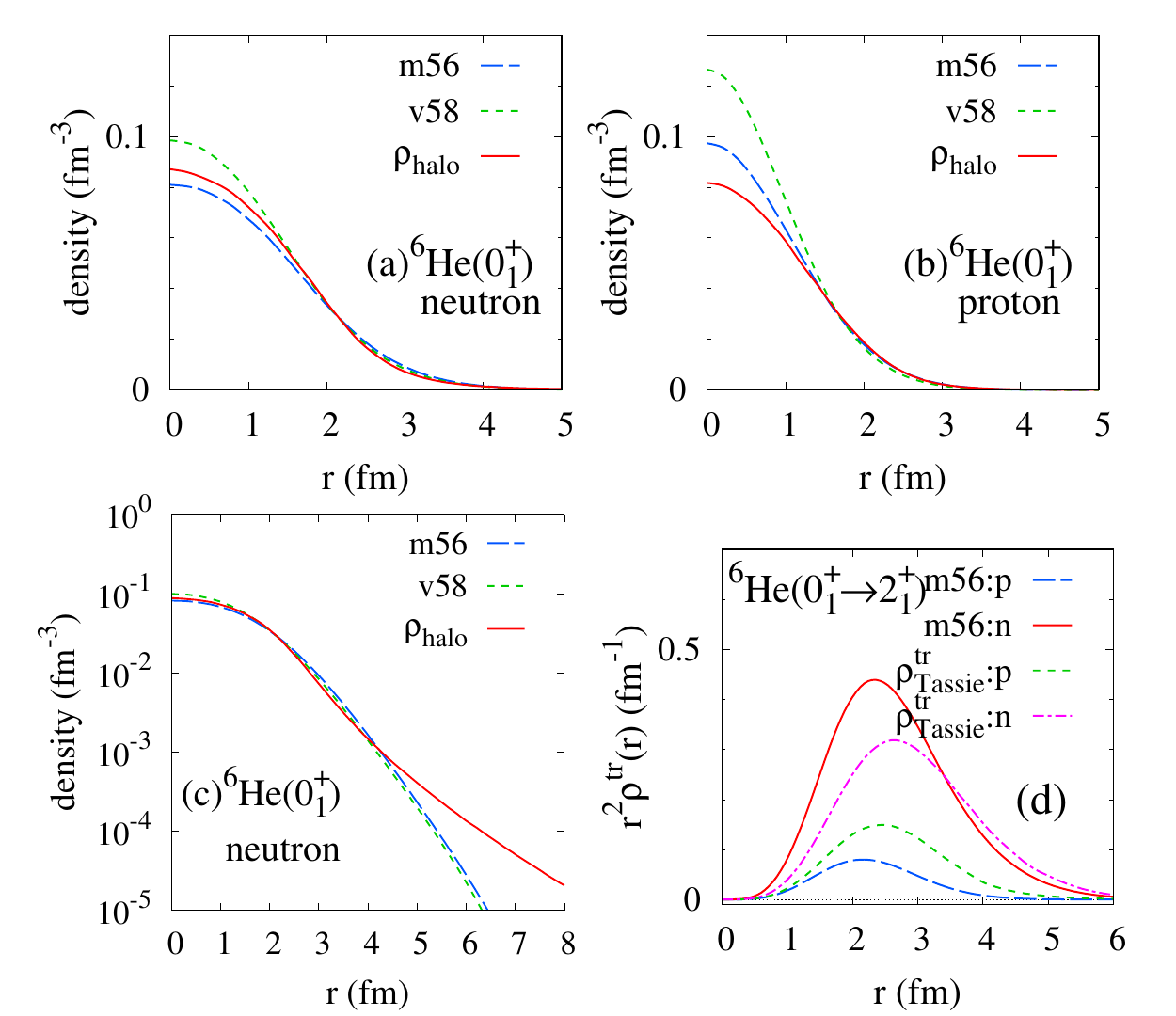}
  \caption{Comparison of the densities of $\Hes$ between 
the halo-type~\cite{Lapoux:2015jva} and AMD densities. 
Panels (a) and (b) compare the neutron and proton components, respectively, 
of the NCSM-based halo-type diagonal density ($\rho_\textrm{halo}$) 
with the AMDm56 and AMDv58 diagonal densities.
Panel (c) shows the neutron densities in log scale. In panel (d), the 2pF-Tassie transition density
($\rho^\textrm{tr}_\textrm{Tassie}$) of the $0^+_1\to 2^+_1$ transition
is compared with the AMDm56 and AMDv58 transition densities. 
The $r^2$-weighted transition densities are also plotted.
The data for $\rho_\textrm{halo}$ are taken from Fig.~13 of Ref.~\cite{Lapoux:2015jva}, and 
those of $\rho^\textrm{tr}_\textrm{Tassie}$ are taken from Fig.~14 of Ref.~\cite{Lapoux:2015jva}.
	\label{fig:trans-he6-halo}}
\end{figure}

\begin{figure}[!htpb]
\includegraphics[width=7 cm]{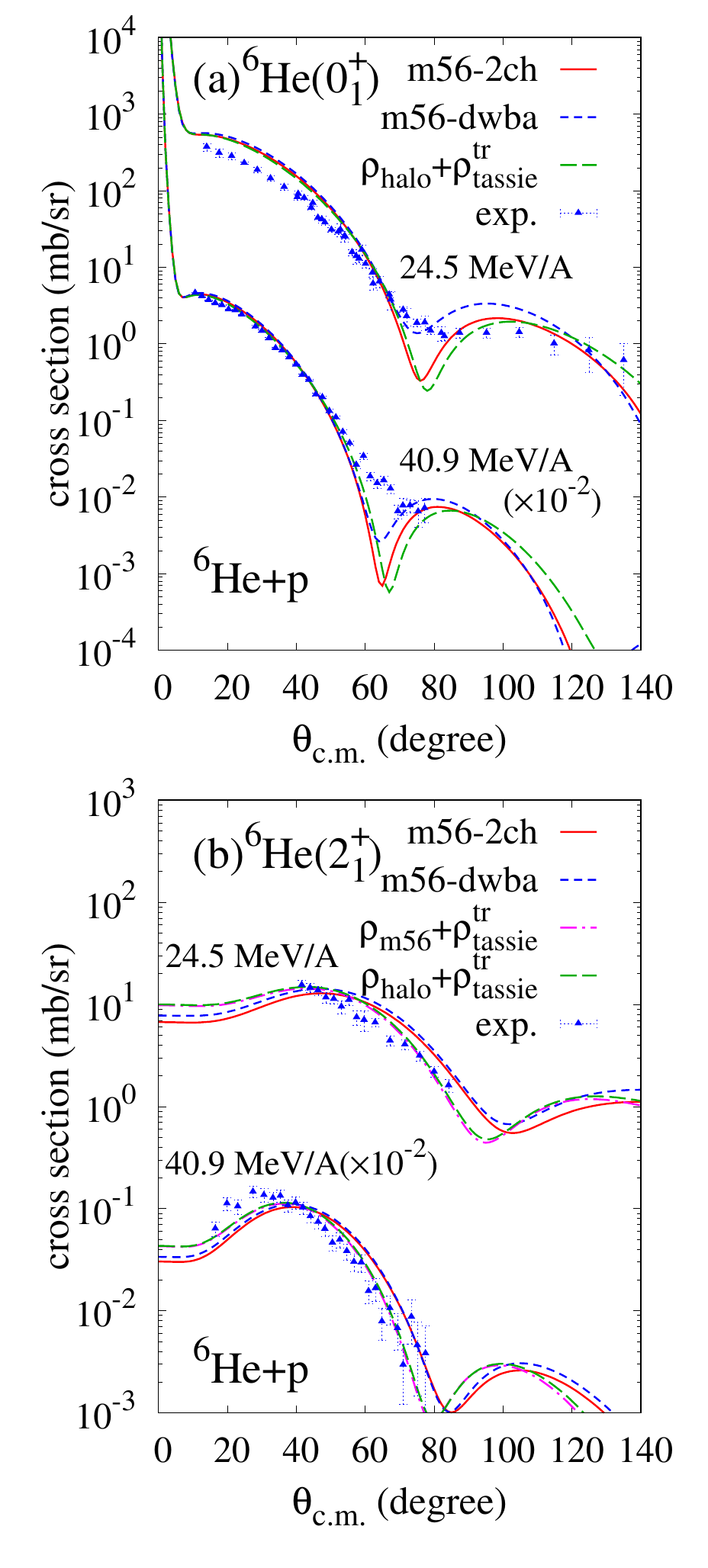}
\caption{
Calculated $\Hes(0^+_1)$ and $\Hes(2^+_1)$
cross sections of the $p+\Hes$ scattering 
at $E=24.5$~MeV/A and $E=40.9$~MeV$~(\times 10^{-2})$
obtained with the AMDm56 densities and those obtained with the halo-type density set.
The 2ch(MCC) and DWBA calculations with AMDm56, and the 2ch(MCC) calculations with 
 $\rho_\textrm{halo}$~(NCSM-based halo-type diagonal density) and 
$\rho^\textrm{tr}_\textrm{Tassie}$~(2pF-Tassie transition density) of $\Hes$ 
from Refs.~\cite{Lagoyannis:2000te,Lapoux:2015jva}. 
The experimental data at $E=24.5$, 25, 25.7~MeV/A~\cite{Stepantsov:2002efb,Giot:2005zz,Wolski:1999ppw} and the data 
at $E=40.9$~MeV/A~$(\times 10^{-2})$~\cite{Lagoyannis:2000te} 
are also shown.
	\label{fig:cross-he6p-halo}}
\end{figure}

\subsection{Ambiguity of structure inputs: comparison with halo-type density}
As previously mentioned, 
the present MCC+AMD results on the $p+\Hes$ reaction were not satisfactory in their detailed description of 
the angle dependence of the $\Hes(2^+_1)$ cross sections.
As can be seen in Fig.~\ref{fig:cross-he6p}(b), the calculation obtained a wider angle distribution than the 
observed cross sections at both energies 
of $E=24.5$ and 40.9~MeV/A.
This energy-independent 
trend may suggest that a modification of the AMD transition densities is 
necessary.

To investigate the ambiguity of the structure inputs, we performed a model analysis 
of the $p+\Hes$ cross sections by adopting the  
halo-type density set of $\rho_\textrm{halo}$ and 
$\rho^\textrm{tr}_\textrm{Tassie}$ presented in Ref.~\cite{Lapoux:2015jva}.
Note that $\rho^\textrm{tr}_\textrm{Tassie}$ is not a microscopic density 
but  a phenomenological transition density given by a derivative form of $\rho_\textrm{halo}$, which was 
renormalized to reproduce the $(p,p')$ data by a JLM reaction calculation 
in Ref.~\cite{Lapoux:2015jva}. 
The proton component of $\rho^\textrm{tr}_\textrm{Tassie}$
was renormalized to fit $B(E2)=3.1\pm 0.6$~$e^2$fm$^4$,
whereas the neutron component was tuned to reproduce 
the $(p,p')$ data. 
Figs.~\ref{fig:trans-he6-halo}(a)-(c) compare the halo-type diagonal density~($\rho_\textrm{halo}$)
with the AMDm56 and AMDv58 diagonal densities of $\Hes$.
Compared with the AMDm56 and AMDv58 densities, 
the neutron component of $\rho_\textrm{halo}$ showed a long tail 
in the $r\ge 5$ fm region (Fig.~\ref{fig:trans-he6-halo}(c)). 
In Fig.~\ref{fig:trans-he6-halo}(d), a comparison of the halo-type transition density~($\rho^\textrm{tr}_\textrm{Tassie}$)
with the AMDm56 transition density of $\Hes$ is shown.
The neutron component of $\rho^\textrm{tr}_\textrm{Tassie}$ shows a broader radial distribution 
compared with AMDm56 but gives almost the same neutron transition matrix element $M_n=7.8$ fm$^2$
as AMDm56 ($M_n=7.9$ fm$^2$).

To see the sensitivity of the $p$ scattering cross sections to the $\Hes$ densities, 
we calculated the $p+\Hes$ reaction using the Melbourne $g$-matrix folding model with the halo-type densities.
Fig.~\ref{fig:cross-he6p-halo} shows a comparison of the results of the $\Hes(0^+_1)$ and $\Hes(2^+_1)$ cross sections 
between two sets of diagonal and transition densities, 
the halo-type density set 
($\rho_\textrm{halo}$ and $\rho^\textrm{tr}_\textrm{Tassie}$) and the AMDm56 density set. Note that the $(p,p)$ cross sections 
were dominantly affected by the diagonal density of the $\Hes(0^+_1)$ state, 
whereas the $(p,p')$ cross sections to the
$\Hes(2^+_1)$ state were sensitive to the
$0^+_1\to 2^+_1$ transition density.
As can be seen in Fig.~\ref{fig:cross-he6p-halo}(a), 
there was only a small difference in the $(p,p)$ cross sections 
between the halo-type and AMDm56 density cases even though they had
different tail behaviors in the neutron diagonal density. 
In the $(p,p')$ cross sections to the $\Hes(2^+_1)$ state, the absolute amplitude of 
the first peak was almost the same as each of the others, though the two calculations resulted in 
different angle dependences.
In the halo-type density case, 
the peak amplitude of the $\Hes(2^+_1)$  cross sections shifted to forward angles and 
the angle distribution was narrower than in the AMDm56 case
because of the slightly broader neutron transition density (Fig.~\ref{fig:trans-he6-halo}~(d)).

From this model analysis, it can be concluded that 
the angle dependence of the $2^+_1$ cross sections was directly affected 
by the radial behavior of the neutron transition density. At the same time, 
the absolute amplitude of the first-peak cross sections was not sensitive to the 
detailed shape of the transition density but was determined by the transition matrix elements.
The results obtained using the halo-type density set demonstrated 
better agreement with the $(p,p')$ data, which may suggest 
that a broader neutron transition density than that of AMDm56 was favored. 
However, there remains 
some deviation from the data for the forward angles at $E=40.9$~MeV/A.
For an accurate description of the $(p,p')$ data observed for the $2^+_1$ resonance energy region,
other effects such as coupling with the continuum states and the $0^+$ and $1^-$ components, have been 
carefully examined by Ogawa and Matsumoto in Ref.~\cite{Ogawa:2020qtt}. 

\section{$p+\Hee$ scattering} \label{sec:he8p}

\subsection{MCC+AMD results for $p+\Hee$ cross sections}

\begin{figure}[!htp]
\includegraphics[width=7 cm]{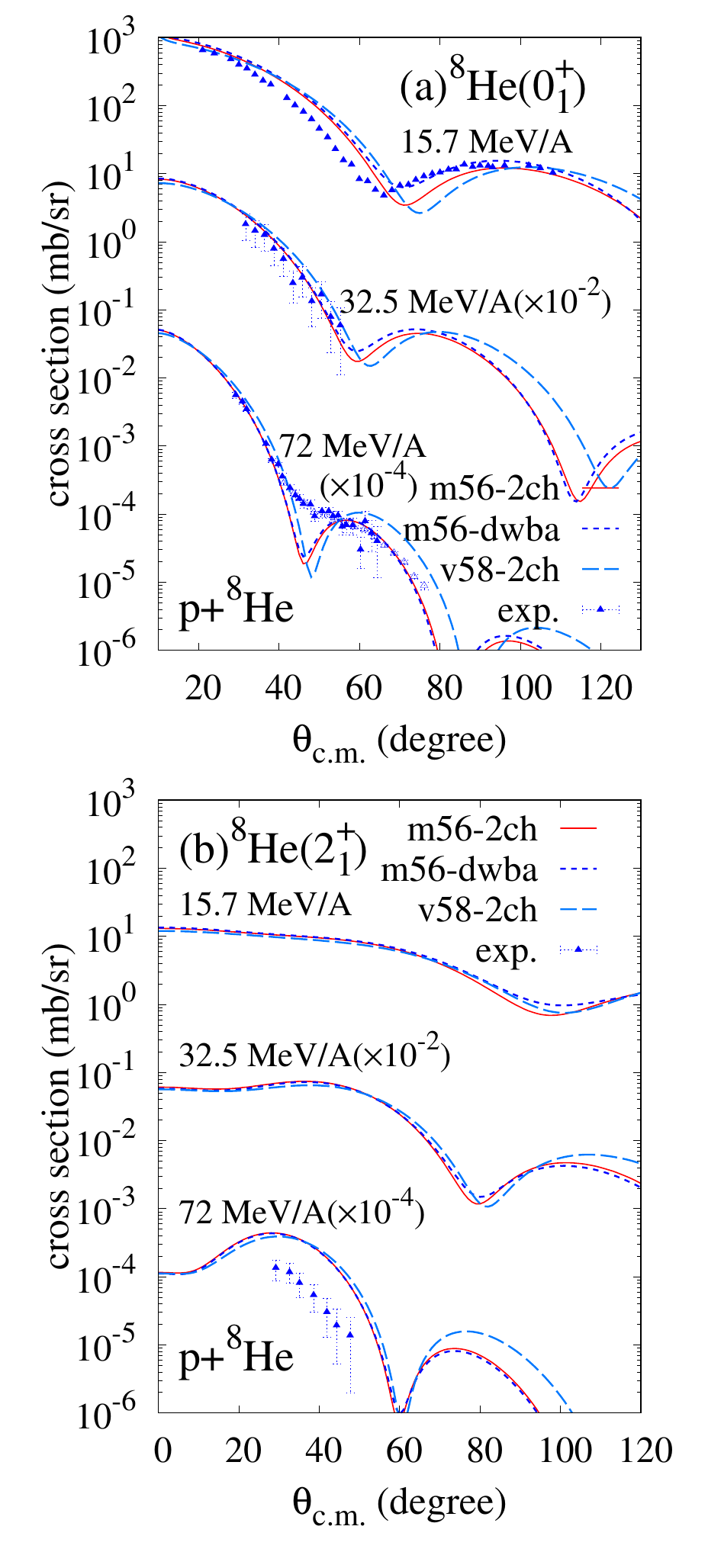}
\caption{
Calculated  $\Hee(0^+_1)$ and  $\Hee(2^+_1)$ cross sections of the $p+\Hee$ scattering 
at $E=15.7$~MeV/A, 32.5~MeV/A ~$(\times 10^{-2})$, and 72~MeV/A~$(\times 10^{-4})$, and the experimental data
from Refs.~\cite{Skaza:2005uff,Korsheninnikov:1995jtx,Korsheninnikov:1993fw}. 
The 2ch(MCC) calculations with AMDm56 and AMDv58 and 
the DWBA calculation with  AMDm56 are shown. 
	\label{fig:cross-he8p}}
\end{figure}

In the same way as the 2ch(MCC) calculation of $p+\Hes$,  
The $p+\Hee$ reaction at
incident energies $E=15.7$, 32.5, and 72 MeV/A was calculated 
by the 2ch(MCC) calculation including the
$\Hee(0^+_1)$ and $\Hee(2^+_1)$ states using the AMDm56 and AMDv58 densities 
as in the 2ch(MCC) calculation of $p+\Hes$.
In addition to the 2ch(MCC) calculation, a DWBA calculation of $p+\Hee$ was performed to obtain  
the one-step cross sections. 

In Fig.~\ref{fig:cross-he8p}(a), 
the calculated $p+\Hee$ elastic cross sections are compared with 
the data observed 
in the inverse kinematics experiments~\cite{Skaza:2005uff,Korsheninnikov:1995jtx,Korsheninnikov:1993fw}.
It should again be noted that the deep dip structure of the present result could be smeared 
by the spin-orbit interaction, which was omitted in the calculation. 
Compared with AMDv58, 
the MCC calculation with AMDm56 obtained a better result for the $(p,p)$ cross sections at $E=$72~MeV/A
for the first- and second-peak amplitudes of the data because of the remarkable neutron skin structure. 
It also successfully reproduced the data at  $E=32.5$~MeV/A. 
For the very low-energy data at $E=$15.7~MeV/A, 
the AMDm56 result was better than the AMDv58 result, but its reproduction of the data was 
not satisfactory. This can be understood to be an effect of the 
lower reliability of the folding model approach for low-energy scattering.
In  comparison with the DWBA~(one-step) cross sections, the CC effect was negligibly small at  
$E=$72~MeV/A but was significant at $E=15.7$~MeV/A, and the MCC result appeared to 
be worse than the DWBA result  around the second peak. 
This could suggest a weaker $0^+_1$-$2^+_1$ coupling than the AMD prediction, but 
it can not be definitively concluded because the applicability of the present reaction approach 
to such low-energy $p$ scattering has not been well-tested. 

With respect to the $(p,p')$ cross sections to the $\Hee(2^+_1)$ state, which are 
shown in Fig.~\ref{fig:cross-he8p}(b),  
the observed data existed only at $E=72$~MeV/A~\cite{Korsheninnikov:1993fw}. 
The MCC calculations with AMDm56 and AMDv58 
considerably overshot the experimental data by a factor of three around
 the peak position.

\begin{figure}[!htp]
\includegraphics[width=7 cm]{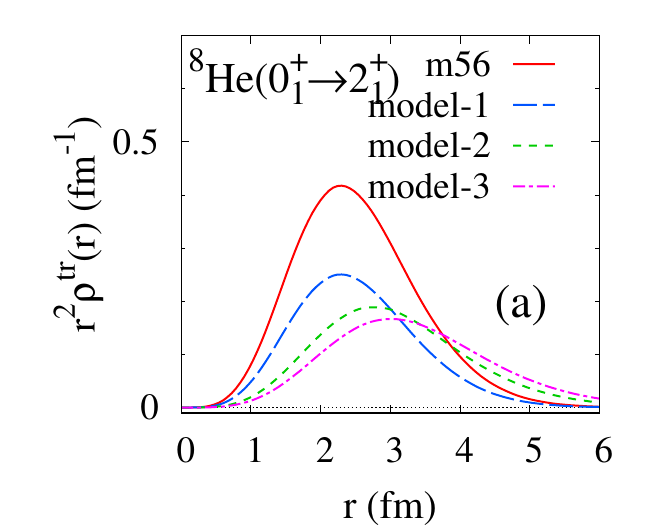}
  \caption{
The neutron transition density of $\Hee$ of 
model functions 
($\rho^\textrm{tr}_{n,\textrm{model-}1}(r)$, $\rho^\textrm{tr}_{n,\textrm{model-}2}(r)$, 
and $\rho^\textrm{tr}_{n,\textrm{model-}3}(r)$) for the $0^+_1\to 2^+_1$ transition
compared with the original AMDm56 density. 
The $r^2$-weighted transition densities are plotted.
The neutron transition matrix element $M_n=7.6$~fm$^2$ for the original AMDm56 density 
was scaled to be 
$M_n=4.5$, 5.3, and 6.1~fm$^2$ for the model-1,  model-2, and  model-3 transition densities, respectively.
	\label{fig:trans-he8-model}}
\end{figure}

\begin{figure}[!htp]
\includegraphics[width=7 cm]{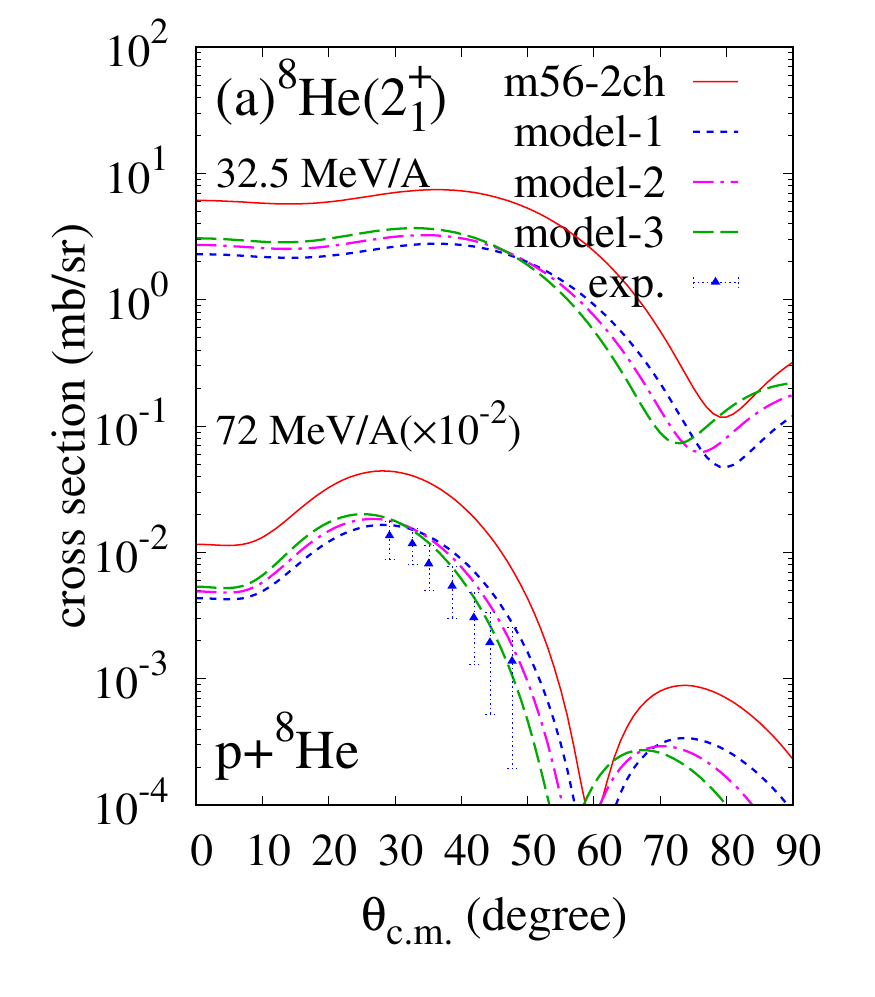}
\caption{Cross sections 
of the $p+\Hee$ scattering 
calculated using the 2ch(MCC) calculations with the model neutron transition densities
of $\rho^\textrm{tr}_{n,\textrm{model-}1}(r)$, $\rho^\textrm{tr}_{n,\textrm{model-}2}(r)$, 
and $\rho^\textrm{tr}_{n,\textrm{model-}3}(r)$.
The calculated $\Hee(2^+_1)$ cross sections 
at $E=32.5$~MeV/A and 72~MeV/A~$(\times 10^{-2})$ 
are compared with the original AMDm56 result and the experimental data from Ref.~\cite{Korsheninnikov:1993fw}. 
	\label{fig:cross-he8p-model}}
\end{figure}

\subsection{Model analysis of $\Hee(2^+)$ cross sections}

As previously described, the present MCC+AMD calculation overshot the experimental $2^+_1$ 
cross sections of the $p+\Hee$ reaction at 
$E=72$~MeV/A by a factor of three. 
A similar overshooting problem for $\Hee(2^+_1)$ cross sections was encountered in the JLM
reaction calculation using the NCSM density of $\Hee$ in Ref.~\cite{Lapoux:2015jva}, 
in which the results were overestimated by a factor of five.
Even though the data at $E=72$~MeV/A contained sizable errors, 
a significant modification of the predicted 
neutron transition density could be used to correct the description of the $(p,p')$ data. 
 
We here consider modifications of the theoretical neutron 
transition density by hand and perform a model analysis to adjust the $(p,p')$ data.
For the modifications, we introduced a model function
$\rho^\textrm{tr}_{n,\textrm{model}}(r)$ for the neutron transition density 
by artificially scaling the original AMDm56 density $\rho^\textrm{tr}_{n,m56}(r)$
as follows:
\begin{equation}
\rho^\textrm{tr}_{n,\textrm{model}}(r)\equiv \frac{f}{a^5}\rho^\textrm{tr}_{n,m56}(r/a).
\end{equation}
Here, $a$ is the radial scaling factor and $f$ is the overall scaling factor.
For this model function, the neutron transition matrix element $M_{n,\textrm{m56}}=7.6$ fm$^2$
of the original AMDm56 value was scaled to be 
$M_n= f M_{n,\textrm{m56}}$.
For the present analysis, we prepared three types of model transition density, $\rho^\textrm{tr}_{n,\textrm{model-1,2,3}}(r)$. 
Model-1 was a transition density 
$\rho^\textrm{tr}_{n,\textrm{model-1}}(r)=f \rho^\textrm{tr}_{n,m56}(r)$ simply renormalized 
from the original transition density using $a=1$ (no radial scaling).
Model-2 and model-3 were obtained by stretching the spatial distribution with radial scaling factors 
of $a=1.1$ and $a=1.2$, respectively. 
We chose the overall factors $f_1=0.6$, $f_2=0.7$, and $f_3=0.8$
for model-1, model-2, and model-3, respectively, to roughly 
fit the upper limit of the $(p,p')$ data at $\theta= 32^\circ$.
As a result, 
these phenomenological model transition densities $\rho^\textrm{tr}_{n,\textrm{model-1}}(r)$, 
$\rho^\textrm{tr}_{n,\textrm{model-3}}(r)$, and $\rho^\textrm{tr}_{n,\textrm{model-3}}(r)$ 
gave the values $M_n=4.5$, 5.3, and 6.1~fm$^2$, respectively, 
which can be regarded as upper limits to reproduce the $(p,p')$ data in each model. 
Figure~\ref{fig:trans-he8-model} shows the neutron transition density of three models. 

Using the model neutron transition densities $\rho^\textrm{tr}_{n,\textrm{model-}1}(r)$, $\rho^\textrm{tr}_{n,\textrm{model-}2}(r)$, and $\rho^\textrm{tr}_{n,\textrm{model-}3}(r)$ of the $0^+_1\to 2^+_1$ transition, 
we performed the coupled-channel calculation of the $p+\Hee$ reaction using
the Melbourne $g$-matrix folding approach. 
We did not change other inputs from the original AMDm56 densities, 
such as the diagonal and 
$2^+_1\to 2^+_1$ transition densities and the proton component of the $0^+_1\to 2^+_1$ transition density.

Figure~\ref{fig:cross-he8p-model} shows the $(p,p')$ cross sections to the $\Hee(2^+_1)$ state 
at $E=32.5$ and 72~MeV/A obtained using the three models
compared with the experimental data and the 
original AMDm56 cross sections.
As mentioned above, the transition density was renormalized by an overall factor
to reproduce the upper limit of the experimental cross sections. 
For the angle dependence of the $(p,p')$ cross sections,
the three models each gave different results. Compared with the experimental $(p,p')$ data, 
the agreement seems to have been better in the model-2 and model-3 results
compared with the model-1 results. This means that 
a distribution of the neutron transition density broader than the original one was favored. 
If we renormalized the neutron transition density of model-2 and model-3 to fit 
the lower limit of the $(p,p')$ data at $\theta= 32^\circ$, we obtained the
neutron transition matrix elements of 3.9 and 4.5~fm$^2$, respectively, 
instead of the values 5.3~(model-2) and 6.1~fm$^2$(model-3) for the upper limit.
Considering these uncertainties, the optimal neutron matrix element to
describe the $(p,p')$ data at $E=72$~MeV/A
 is likely to be in the range of $M_n=$4--6~fm$^2$.

\section{$p+\Hee$ and $\alpha+\Hee$ inelastic scattering}
\label{sec:he8-J0123}

\begin{figure}[!htpb]
\includegraphics[width=9 cm]{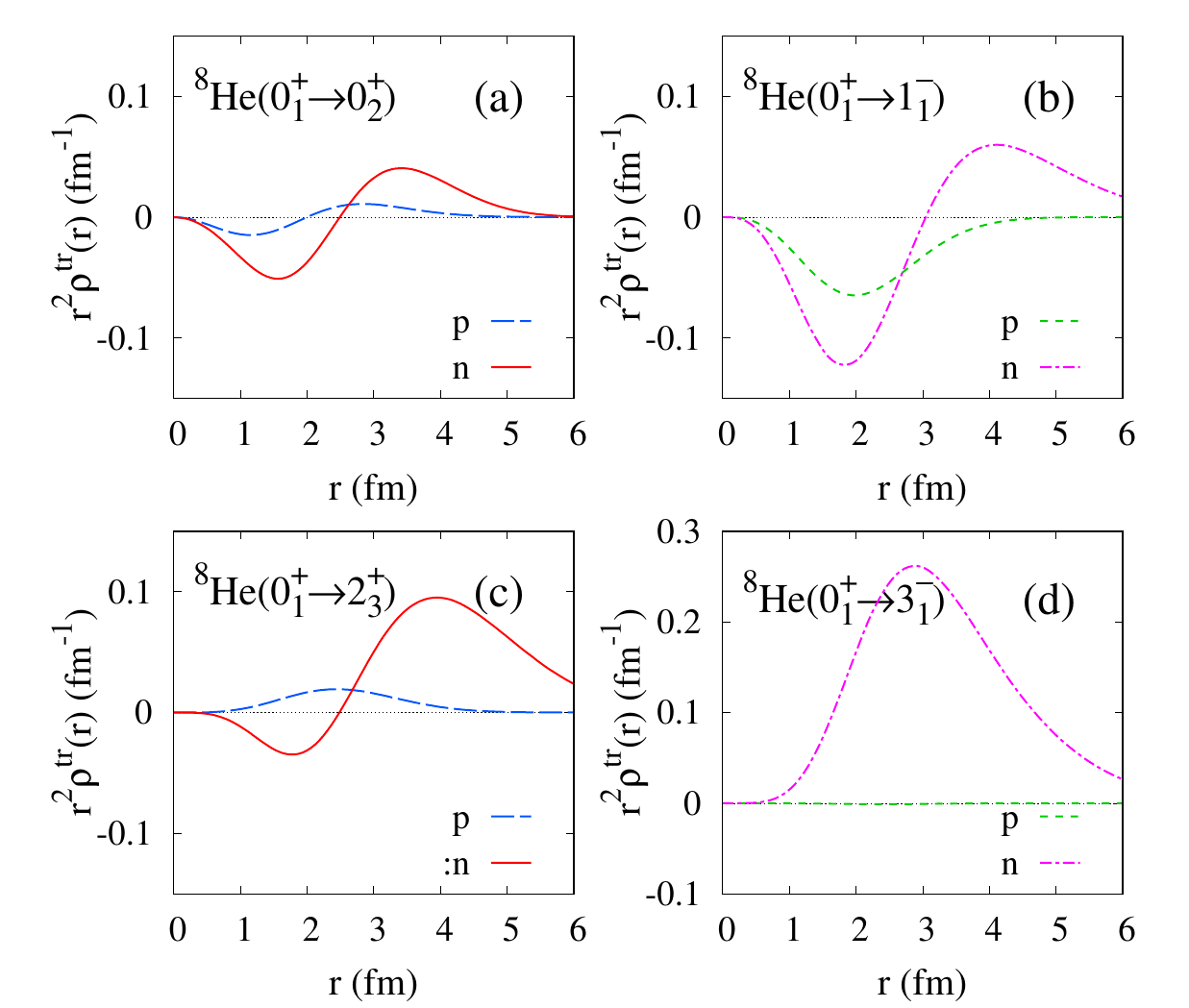}
  \caption{Proton and neutron transition densities from the $\Hee(0^+_1)$ 
state to the $\Hee(0^+_2)$, $\Hee(1^-_1)$, $\Hee(2^+_3)$ and $\Hee(3^-_1)$ states.
	\label{fig:trans-he8-J0123}}
\end{figure}

To investigate the $p$ and $\alpha$ inelastic scattering off $\Hee$, 
we performed MCC+AMD calculations of the $p+\Hee$ and $\alpha+\Hee$ reactions. 
These calculations included the $0^+_{1,2}$, $1^-_1$, $2^+_{1,2,3}$, and $3^-_1$ states (seven states 
in total; called ``7ch(MCC)'')  
and all the $\lambda=0,1,2,3$ 
transitions with the AMDm56 densities. 
As seen in the calculated $M_n/M_p$ ratios shown in Table~\ref{tab:he8-J0123}, 
the inelastic transitions from the
$0^+_1$ state to the $0^+_2$, $1^-_1$, $2^+_3$, and $3^-_1$ states had a remarkable neutron-dominant nature.
The proton and neutron components of the AMDm56 transition densities
are shown in Fig.~\ref{fig:trans-he8-J0123}. 
The neutron component of the $0^+_1\to 0^+_2$ and $0^+_1\to 1^-_1$ transition densities 
showed nodal structures as are usually seen in the isoscalar monopole and 
dipole transitions of $Z=N$ nuclei. The $0^+_1\to 3^-_1$ neutron transition density 
had a single-peak structure, whereas 
the $0^+_1\to 2^+_3$ neutron transition density had a nodal structure 
different from the $0^+_1\to 2^+_1$ transition with the simple peak.

\begin{figure}[!htp]
\includegraphics[width=7 cm]{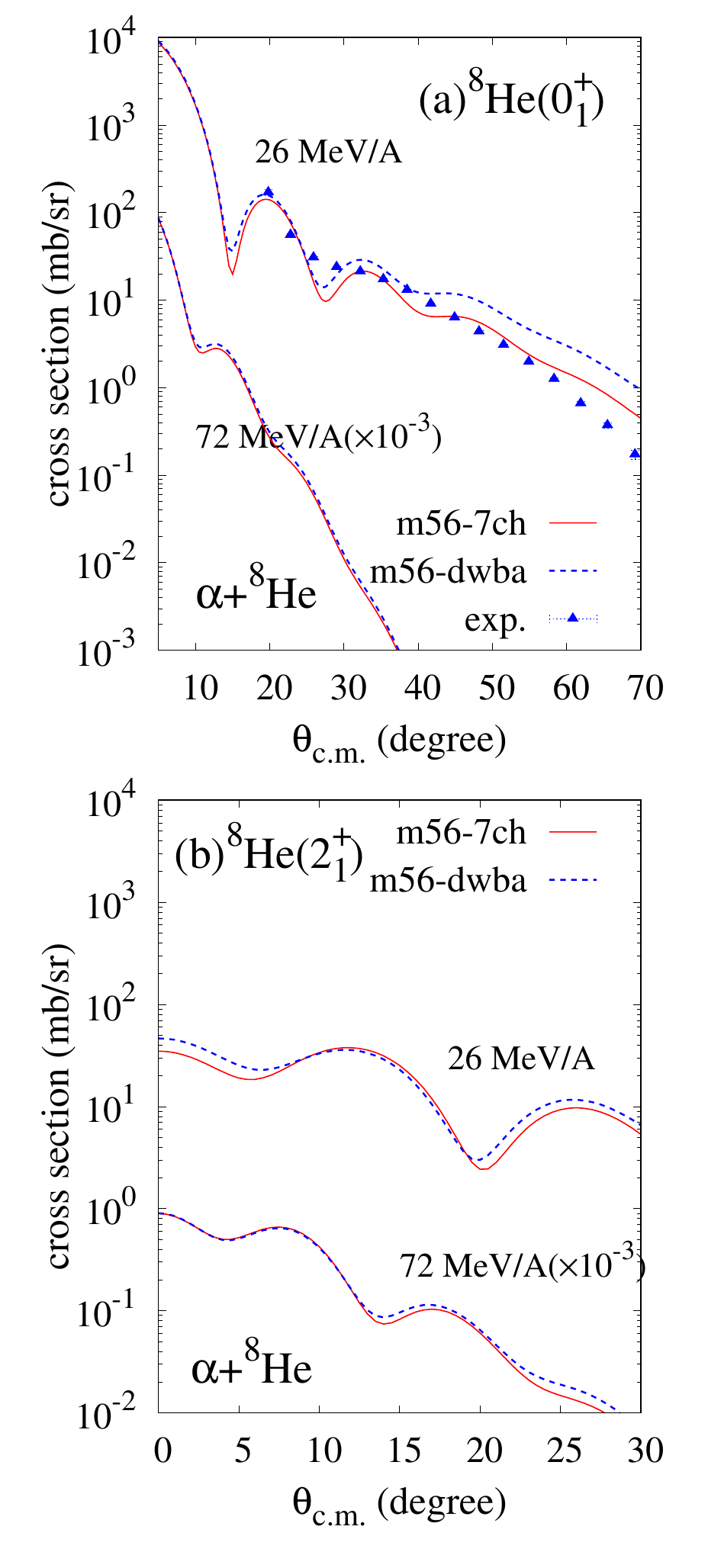}
\caption{ Calculated $\Hee(0^+_1)$ and $\Hee(2^+_1)$ cross sections of the 
$\alpha+\Hee$ reaction at $E=26$~MeV/A and 72~MeV/A~$(\times 10^{-2})$
obtained by the 7ch(MCC) and DWBA calculations with the AMDm56 densities. 
The experimental data of the elastic cross sections at  $E=26$~MeV/A from 
Refs.~\cite{Wolski:2002gzz,Wolski:2003moy} are also shown.
	\label{fig:cross-he8a}}
\end{figure}

\begin{figure}[!htp]
\includegraphics[width=7 cm]{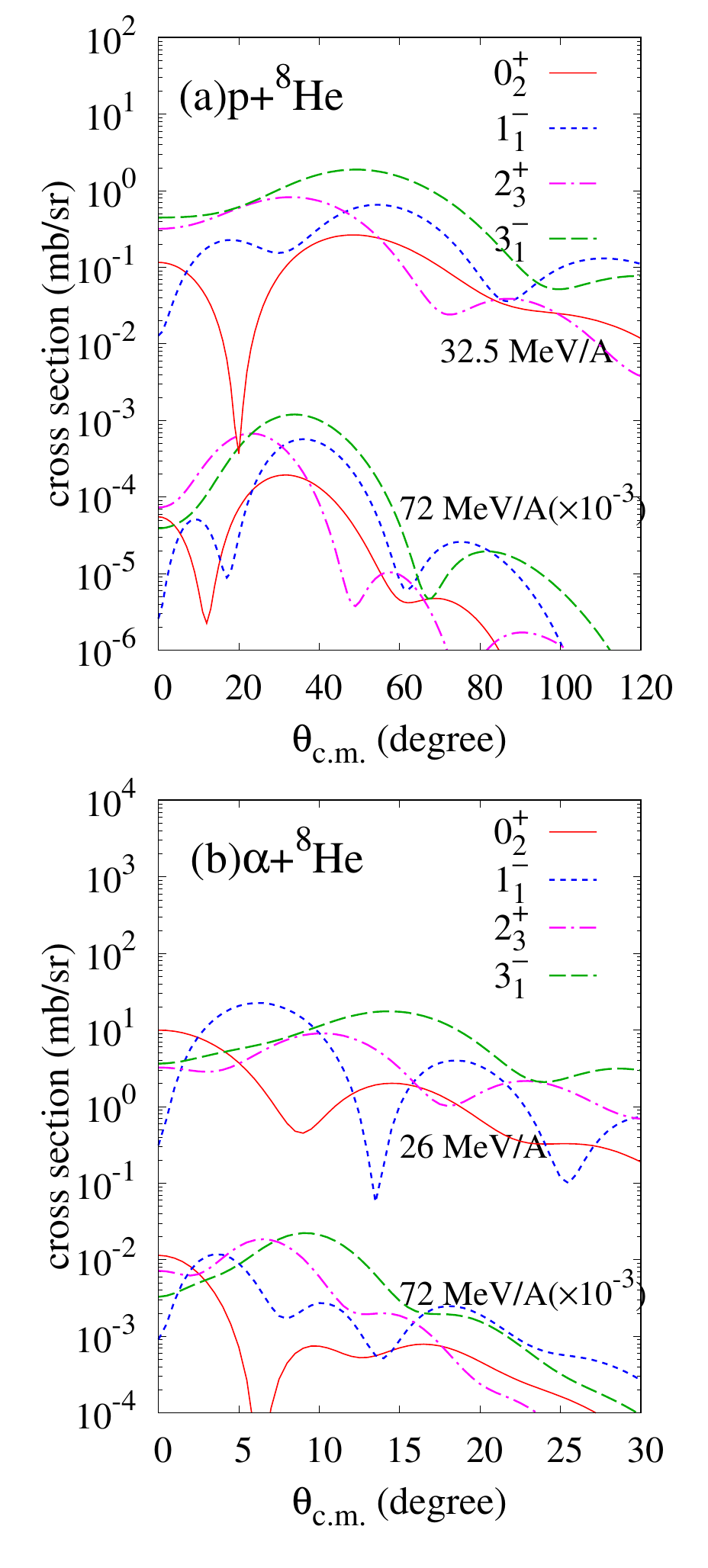}
\caption{Calculated $(p,p')$ cross sections at $E=32.5$~MeV/A and 72~MeV/A$(\times 10^{-2})$ and 
($\alpha,\alpha')$ cross sections at $E=26$~MeV/A and 72~MeV/A$(\times 10^{-2})$ to the 
$\Hee(0^+_2)$, $\Hee(1^-_1)$, $\Hee(2^+_3)$, and $\Hee(3^-_1)$ states
obtained by the 7ch(MCC) calculations with the AMDm56 densities.  
	\label{fig:cross-he8-J0123}}
\end{figure}

\begin{figure}[!htp]
\includegraphics[width=7 cm]{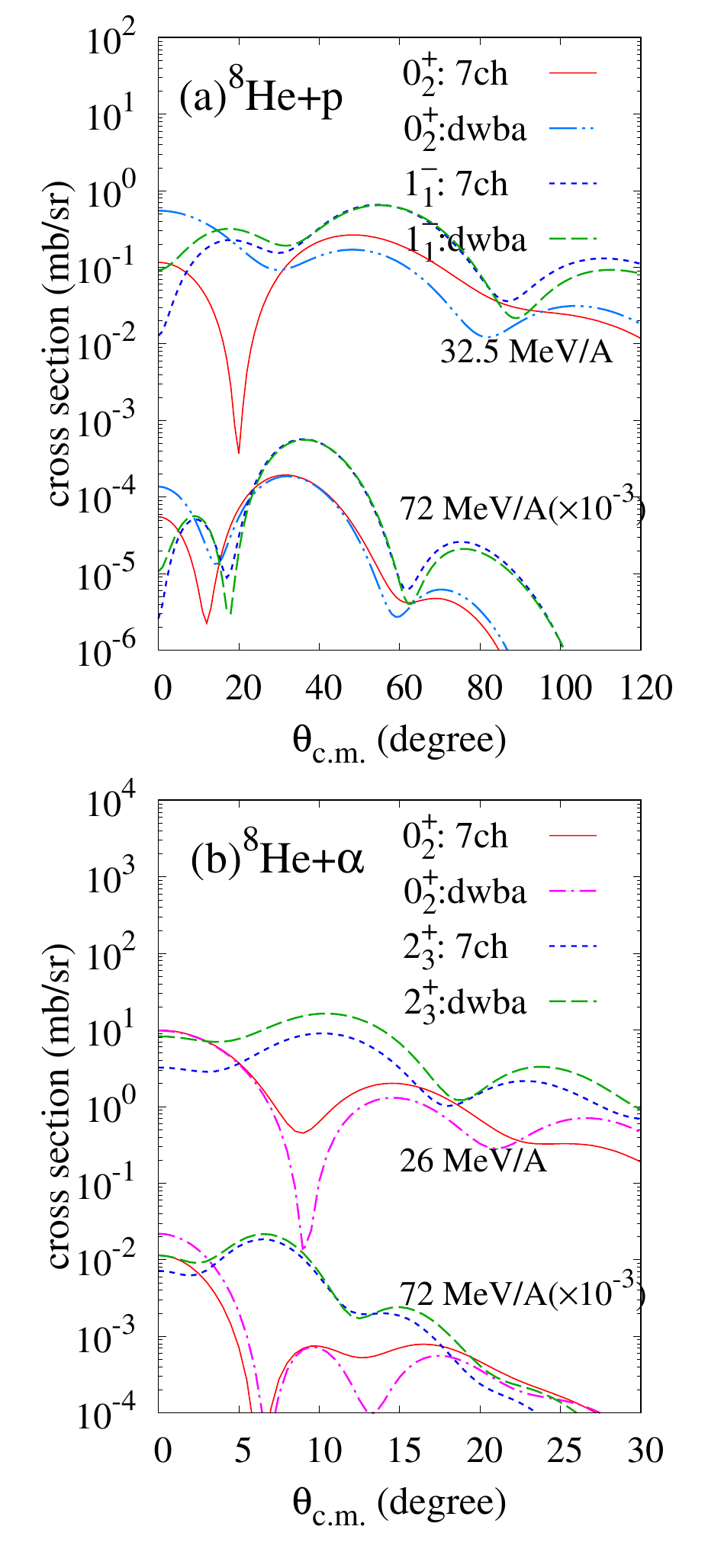}
\caption{DWBA cross sections of the 
$p+\Hee$ reaction at $E=32.5$~MeV/A and 72~MeV/A$(\times 10^{-2})$ and of the $\alpha+\Hee$ reaction 
at $E=26$~MeV/A and 72~MeV/A$(\times 10^{-2})$ to the $\Hee(0^+_2)$ and  $\Hee(1^-_1)$ states
compared with the 7ch(MCC) calculations.
	\label{fig:cross-he8-J0123-dwba}}
\end{figure}

Figure~\ref{fig:cross-he8a} shows the 
elastic and $2^+_1$ cross sections of the 
$\alpha+\Hee$ reaction at $E=26$ and 72~MeV/A 
obtained using the 7ch(MCC) calculation with the AMDm56 densities
together with the experimental elastic cross sections at $E=26$~MeV/A~\cite{Wolski:2002gzz,Wolski:2003moy}. 
The MCC calculation successfully reproduced the observed $(\alpha,\alpha)$ data.
In comparison with the DWBA~(one-step) cross sections,
the CC effect had a minor contribution to the $\Hee(0^+_1)$ and $\Hee(2^+_1)$ cross sections
of the $\alpha+\Hee$ scattering except for in the case of elastic scattering at backward angles.

The MCC+AMD results of the $0^+_2$, $1^-_1$, $2^+_3$, and $3^-_1$ cross sections 
of the $p+\Hee$ and $\alpha+\Hee$ reactions are shown in Fig.~\ref{fig:cross-he8-J0123}.
According to the AMDm56 prediction, the $\Hee(0^+_2)$, $\Hee(1^-_1)$, and $\Hee(3^-_1)$ states had 
few MeV energy differences.
For experimental search for these states with $(p,p')$ and $(\alpha,\alpha')$ reactions,
the production rate and selectivity of each state should be carefully considered.
In particular, the $(p,p')$ reaction may be favored to observe the $\Hee(3^-_1)$ state because of its high production 
rate for a wide range of angles. At the same time, the $(\alpha,\alpha')$ reaction 
had a high selectivity for the $\Hee(0^+_2)$ state at forward 
angles  ($\theta_\textrm{c.m.}\le 2^\circ$). 
For the  $\Hee(1^-_1)$ state, the dominant contribution was predicted in the $(\alpha,\alpha')$ cross sections
around $\theta_\textrm{c.m.}\sim 5^\circ$ and at a lower energy of $E=26$~MeV/A. 

To see the CC effect in the $p+\Hee$ and $\alpha+\Hee$ reactions, 
the DWBA~(one-step) cross sections were compared with the MCC calculations
 (Fig.~\ref{fig:cross-he8-J0123-dwba}). 
Because of the strong $0^+_2\to 2^+_3$ coupling, 
a significant CC effect can be observed 
in the $(p,p')$ cross sections to the $\Hee(0^+_2)$ state and the $(\alpha,\alpha')$ cross sections
to the $\Hee(0^+_2)$ and $\Hee(2^+_3)$ states, 
for the low energy reactions in particular.

The relative high productions of the $\Hee(0^+_2)$ and $\Hee(1^-_1)$ states in the $(\alpha,\alpha')$ reaction 
compared with the $(p,p')$ reaction can be understood in terms of the nodal behavior of the transition density.
In general, the $(\alpha,\alpha')$ reaction is 
a sensitive probe to the transition density at the surface and outer regions 
because of the strong absorption of the $\alpha$$-$nucleus potentials in the interior region. 
As mentioned earlier
the neutron transition densities from the ground to the $\Hee(0^+_2)$ and $\Hee(1^-_1)$ states 
had a remarkable amplitude in the outer region (Figs.~\ref{fig:trans-he8-J0123}(a) and (b)), 
and this outer amplitude predominantly contributed to the $(\alpha,\alpha')$ cross sections. 
At the same time, in the $p$ scattering with a weaker absorption, 
the inner amplitude had a negative contribution and canceled out the
contribution of the outer amplitude and suppressed the forward angle cross sections. 
High productions of the monopole and dipole transitions by $(\alpha,\alpha')$ reactions
have been observed for isoscalar transitions in $Z\eqsim N$ nuclei. The new finding obtained 
here is that
a similar trend was also predicted  
in neutron-dominant transitions in neutron-rich nuclei.

\section{Summary} \label{sec:summary}

The elastic and inelastic cross sections of the $p+\Hes$, $p+\Hee$, and $\alpha+\Hee$ reactions 
were investigated using the Melbourne $g$-matrix folding approach. 
In the reaction calculations, the transition and coupling potentials of nucleon$-$nucleus systems 
were microscopically obtained by folding the Melbourne $g$-matrix $NN$ interaction with the diagonal 
and transition densities of the target nucleus, and the 
$\alpha$$-$nucleus potentials were calculated by folding the nucleon$-$nucleus potentials with 
the $\alpha$ density. 
The theoretical densities obtained using the microscopic structure model of AMD were used 
for the MCC calculations of the $p+\Hes$, $p+\Hee$, and $\alpha+\Hee$ reactions.
One of the unique characteristics of the present work using MCC+AMD
is that we investigated the $p+\Hes$, $p+\Hee$, and $\alpha+\Hee$ reactions 
on the same footing in the microscopic framework. We first demonstrated the success of the MCC+AMD calculation 
in reproducing elastic cross sections, and then discussed the transition properties of the $\Hes(2^+_1)$ 
and $\Hee(2^+_1)$ states 
with detailed analyses of the $(p,p')$ cross sections.
It should be stressed that the Melbourne $g$-matrix folding approach has no adjustable parameters
differently from phenomenological reaction analysis.
This is a great advantage to test the reliability of 
structure model inputs via the cross section data. 

For the $p+\Hes$ and $p+\Hee$ elastic scattering, 
the present MCC+AMD calculation reproduced well the  $p+\Hes$  cross sections 
at E=40.9~MeV/A  and the $p+\Hee$ cross sections at $E=32.5$ and 72~MeV/A.  

For the $p+\Hes$ inelastic scattering to the $\Hes(2^+_1)$ state at $E=24.5$, and 40.9~MeV/A, 
the MCC+AMD results demonstrated a reasonable reproduction of the peak amplitude, supporting
the predicted value of $M_n=7.9$ fm$^2$. 
However, the result was not satisfactory in describing 
the angle dependence of the $(p,p')$ data in detail. 
We performed a model analysis using the Melbourne $g$-matrix folding approach with 
phenomenological halo-type densities and demonstrated that a better reproduction can be 
obtained using a broader neutron transition density 
than that in the AMD prediction. 

For the $\Hee(2^+_1)$ cross sections of the $p+\Hee$ reaction,
the present MCC+AMD calculation considerably overshot the $(p,p')$ data at $E=72$~MeV/A by
a factor of three. This overshooting was 
consistent with the reaction analysis with the NCSM density performed in Ref.~\cite{Lapoux:2015jva}. 
To gather information concerning the neutron transition from the $(p,p')$ data, 
we performed a further model analysis by introducing
phenomenological modifications of the neutron component of the $0^+_1\to 2^+_1$ transition density
to fit the $(p,p')$ data at $E=72$~MeV/A, 
and obtained a plausible value of $M_n=$4--6~fm$^2$, which was smaller 
than that obtained by the AMD predictions ($M_n=6.4$--7.6~fm$^2$). 
For a more detailed discussion on this point, 
high-quality data for a wide energy range would be  required. 

For $p$ scattering below $E=25$~MeV/A, the present results were not satisfactory 
in precisely reproducing the observed $p+\Hes$ and $p+\Hee$ elastic cross sections. 
The applicability of the present reaction approach to such the 
low-energy $p$ scattering should be carefully examined. 
For example, the validity of the local density approximation treatment 
for light-mass nuclei with large isospin asymmetry remains to be checked.
Coupling with continuum states may also contribute to the low-energy $p$ scattering 
off loosely bound nuclei. 

For the $\alpha+\Hee$ reaction, the MCC+AMD calculation
reproduced the observed $(\alpha,\alpha)$ cross sections at $E=26$~MeV/A. 
The theoretical predictions of the $(p,p')$ and $(\alpha,\alpha')$ cross sections to 
higher excited states of 
$\Hee$ were presented, and the production rates in the $p+\Hee$ and $\alpha+\Hee$ inelastic scattering
were discussed. 
It was suggested that the $(p,p')$ reaction was
favored for the $\Hee(3^-_1)$ observation, whereas the $(\alpha,\alpha')$ reaction at forward angles 
sensitively probed the $\Hee(0^+_2)$ and $\Hee(1^-_1)$ states. 
Our results indicate that $\alpha$ inelastic scattering has the potential to be a good probe 
for neutron-dominant monopole and dipole excitations, and its
wide application to neutron-rich nuclei is expected in future experiments in inverse kinematics. 

\begin{acknowledgments}
The authors would like to thank Dr.~Matsumoto and Mr.~Ogawa for their fruitful discussions. 
The computational calculations of this work were performed using the
supercomputer at the Yukawa Institute for Theoretical Physics at Kyoto University. The work was supported
by Grants-in-Aid of the Japan Society for the Promotion of Science (Grant Nos. JP18K03617, JP16K05352, and 18H05407) and by a grant of  the joint research project of the Research Center for Nuclear Physics at Osaka University.
\end{acknowledgments}


\begin{thebibliography}{9}
\bibitem{Harakeh-textbook}
M.N.~Harakeh, A.~van der Woude, Giant Resonances, Oxford University Press, 2001.

\bibitem{VanDerBorg:1981qiu} 
  K.~Van Der Borg, M.~N.~Harakeh and A.~Van Der Woude,
  Nucl.\ Phys.\ A {\bf 365}, 243 (1981).

\bibitem{Youngblood:1999zza} 
  D.~H.~Youngblood, H.~L.~Clark and Y.-W.~Lui,
  Phys.\ Rev.\ Lett.\  {\bf 82}, 691 (1999).

\bibitem{Uchida:2004bs} 
  M.~Uchida {\it et al.},
  Phys.\ Rev.\ C {\bf 69}, 051301 (2004).


\bibitem{Bernstein:1977wtr}
  A.~M.~Bernstein, V.~R.~Brown and V.~A.~Madsen,
  Phys.\ Lett.\  {\bf 71B}, 48 (1977).

\bibitem{Bernstein:1979zza} 
  A.~M.~Bernstein, V.~R.~Brown and V.~A.~Madsen,
  Phys.\ Rev.\ Lett.\  {\bf 42}, 425 (1979).

\bibitem{Bernstein:1981fp} 
  A.~M.~Bernstein, V.~R.~Brown and V.~A.~Madsen,
  Phys.\ Lett.\  {\bf 103B}, 255 (1981).


\bibitem{Brown:1980zzd} 
  B.~A.~Brown and B.~H.~Wildenthal,
  Phys.\ Rev.\ C {\bf 21}, 2107 (1980).


\bibitem{Brown:1982zz}
  B.~A.~Brown {\it et al.},
  Phys.\ Rev.\  C {\bf 26}, 2247 (1982).

\bibitem{Wakasa:2006nt}
  T.~Wakasa {\it et al.},
  Phys.\ Lett.\ B {\bf 653}, 173 (2007).


\bibitem{Itoh:2011zz}
  M.~Itoh {\it et al.},
  Phys.\ Rev.\ C {\bf 84}, 054308 (2011).

\bibitem{Kawabata:2013xea} 
  T.~Kawabata {\it et al.},
  Few Body Syst.\  {\bf 54}, 93 (2013).

\bibitem{Adachi:2018pql}
  S.~Adachi {\it et al.},
  Phys.\ Rev.\ C {\bf 97}, no. 1, 014601 (2018).


\bibitem{Wolski:2002gzz} 
  R.~Wolski {\it et al.},
  Nucl.\ Phys.\ A {\bf 701}, 29 (2002).

\bibitem{Wolski:2003moy} 
  R.~Wolski {\it et al.},
  Nucl.\ Phys.\ A {\bf 722}, C55 (2003).

\bibitem{Furuno:2019lyp} 
  T.~Furuno {\it et al.},
  Phys.\ Rev.\ C {\bf 100}, no. 5, 054322 (2019).

\bibitem{Jeukenne:1977zz} 
  J.~P.~Jeukenne, A.~Lejeune and C.~Mahaux,
  Phys.\ Rev.\ C {\bf 16}, 80 (1977).

\bibitem{Lapoux:2001kpc} 
  V.~Lapoux {\it et al.},
  Phys.\ Lett.\ B {\bf 517}, 18 (2001).

\bibitem{Skaza:2005uff} 
  F.~Skaza {\it et al.},
  Phys.\ Lett.\ B {\bf 619}, 82 (2005).
  doi:10.1016/j.physletb.2005.05.061

\bibitem{Jouanne:2005pb} 
  C.~Jouanne {\it et al.},
  Phys.\ Rev.\ C {\bf 72}, 014308 (2005).

\bibitem{Takashina:2005bs} 
  M.~Takashina, Y.~Kanada-En'yo and Y.~Sakuragi,
  Phys.\ Rev.\ C {\bf 71}, 054602 (2005).

\bibitem{Takashina:2008zza} 
  M.~Takashina and Y.~Kanada-En'yo,
  Phys.\ Rev.\ C {\bf 77}, 014604 (2008).

\bibitem{Lapoux:2015jva} 
  V.~Lapoux and N.~Alamanos,
  Eur.\ Phys.\ J.\ A {\bf 51}, no. 7, 91 (2015).

\bibitem{Matsumoto:2017mau}
T.~Matsumoto, J.~Tanaka and K.~Ogata,
PTEP \textbf{2019}, no.12, 123D02 (2019).

\bibitem{Ogawa:2020qtt}
S.~Ogawa and T.~Matsumoto,
[arXiv:2003.05123 [nucl-th]].

\bibitem{Amos:2000}
K. Amos, P. J. Dortmans, H. V. von Geramb, S. Karataglidis, and J. Raynal,
Adv.~Nucl.~Phys. {\bf 25}, 275 (2000).

\bibitem{Lagoyannis:2000te} 
  A.~Lagoyannis {\it et al.},
  Phys.\ Lett.\ B {\bf 518}, 27 (2001).

\bibitem{Stepantsov:2002efb} 
  S.~V.~Stepantsov {\it et al.},
  Phys.\ Lett.\ B {\bf 542}, 35 (2002).

\bibitem{Karataglidis:2007yj} 
  S.~Karataglidis, Y.~J.~Kim and K.~Amos,
  Nucl.\ Phys.\ A {\bf 793}, 40 (2007).

\bibitem{Minomo:2009ds} 
  K.~Minomo, K.~Ogata, M.~Kohno, Y.~R.~Shimizu and M.~Yahiro,
  J.\ Phys.\ G {\bf 37}, 085011 (2010).

\bibitem{Toyokawa:2013uua} 
  M.~Toyokawa, K.~Minomo and M.~Yahiro,
  Phys.\ Rev.\ C {\bf 88}, no. 5, 054602 (2013).

\bibitem{Egashira:2014zda}
  K.~Egashira, K.~Minomo, M.~Toyokawa, T.~Matsumoto and M.~Yahiro,
  Phys.\ Rev.\ C {\bf 89}, 064611 (2014).

\bibitem{Minomo:2016hgc}
  K.~Minomo and K.~Ogata,
  Phys.\ Rev.\ C {\bf 93}, 051601 (2016).

\bibitem{Minomo:2017hjl} 
  K.~Minomo, K.~Washiyama and K.~Ogata,
  arXiv:1712.10121 [nucl-th].

\bibitem{Kanada-Enyo:2019prr} 
  Y.~Kanada-En'yo and K.~Ogata,
  Phys.\ Rev.\ C {\bf 99}, no. 6, 064601 (2019).

\bibitem{Kanada-Enyo:2019qbp} 
  Y.~Kanada-En'yo and K.~Ogata,
  arXiv:1904.03811 [nucl-th].

\bibitem{Kanada-Enyo:2019uvg} 
  Y.~Kanada-En'yo and K.~Ogata,
  Phys.\ Rev.\ C {\bf 100}, no. 6, 064616 (2019).

\bibitem{Kanada-Enyo:2020zpl} 
  Y.~Kanada-En'yo and K.~Ogata,
  arXiv:2002.02625 [nucl-th].


\bibitem{KanadaEnyo:1995tb}
  Y.~Kanada-En'yo, H.~Horiuchi and A.~Ono,
  Phys.\ Rev.\  C {\bf 52}, 628  (1995).

\bibitem{KanadaEnyo:1995ir}
  Y.~Kanada-En'yo and H.~Horiuchi,
  Phys.\ Rev.\  C {\bf 52}, 647 (1995).


\bibitem{KanadaEn'yo:2012bj}
  Y.~Kanada-En'yo, M.~Kimura and A.~Ono,
  PTEP {\bf 2012}  01A202 (2012).

\bibitem{Ker59}
A. K. Kerman, H. McManus, and R. M. Thaler,
Ann. Phys. {\bf 8}, 551 (1959).



\bibitem{Machleidt:1987hj}
R.~Machleidt, K.~Holinde and C.~Elster,
Phys.\ Rept.\  \textbf{149}, 1-89 (1987).

\bibitem{Kanada-Enyo:2007iri} 
  Y.~Kanada-En'yo,
  Phys.\ Rev.\ C {\bf 76}, 044323 (2007).

\bibitem{Wang:2004ze} 
  L.-B.~Wang {\it et al.},
  Phys.\ Rev.\ Lett.\  {\bf 93}, 142501 (2004).

\bibitem{Mueller:2008bj} 
  P.~Mueller {\it et al.},
  Phys.\ Rev.\ Lett.\  {\bf 99}, 252501 (2007).

\bibitem{Tanihata:1992wf} 
  I.~Tanihata, D.~Hirata, T.~Kobayashi, S.~Shimoura, K.~Sugimoto and H.~Toki,
  Phys.\ Lett.\ B {\bf 289}, 261 (1992).

\bibitem{Ozawa:2001hb} 
  A.~Ozawa, T.~Suzuki and I.~Tanihata,
  Nucl.\ Phys.\ A {\bf 693}, 32 (2001).

\bibitem{Tilley:2002vg} 
  D.~R.~Tilley, C.~M.~Cheves, J.~L.~Godwin, G.~M.~Hale, H.~M.~Hofmann, J.~H.~Kelley, C.~G.~Sheu and H.~R.~Weller,
  Nucl.\ Phys.\ A {\bf 708}, 3 (2002).

\bibitem{Tilley:2004zz} 
  D.~R.~Tilley, J.~H.~Kelley, J.~L.~Godwin, D.~J.~Millener, J.~E.~Purcell, C.~G.~Sheu and H.~R.~Weller,
  Nucl.\ Phys.\ A {\bf 745}, 155 (2004).


\bibitem{Varga:1994fu} 
  K.~Varga, Y.~Suzuki and Y.~Ohbayasi,
  Phys.\ Rev.\ C {\bf 50}, 189 (1994).

\bibitem{Caurier:2005rb} 
  E.~Caurier and P.~Navratil,
  Phys.\ Rev.\ C {\bf 73}, 021302 (2006).

\bibitem{Navratil:2002zz} 
  P.~Navratil and W.~E.~Ormand,
  Phys.\ Rev.\ Lett.\  {\bf 88}, 152502 (2002).

\bibitem{Navratil:2003ef} 
  P.~Navratil and W.~E.~Ormand,
  Phys.\ Rev.\ C {\bf 68}, 034305 (2003).


\bibitem{Giot:2005zz}
L.~Giot, P.~Roussel-Chomaz, C.~Demonchy, W.~Mittig, H.~Savajols, N.~Alamanos, F.~Auger, A.~Gillibert, C.~Jouanne, V.~Lapaoux, L.~Nalpas, E.~Pollacco, J.~Sida, F.~Skaza, M.~Cortina-Gil, J.~Fernandez-Vasquez, R.~Mackintosh, A.~Pakou, S.~Pita, A.~Rodin, S.~Stepantsov, T.~Akopian, G.M., K.~Rusek, I.~Thompson and R.~Wolski,
Phys.\ Rev.\ C \textbf{71}, 064311 (2005).


\bibitem{Wolski:1999ppw} 
  R.~Wolski {\it et al.},
  Phys.\ Lett.\ B {\bf 467}, 8 (1999).

\bibitem{Korsheninnikov:1993fw} 
  A.~A.~Korsheninnikov {\it et al.},
  Phys.\ Lett.\ B {\bf 316}, 38 (1993).

\bibitem{Korsheninnikov:1995jtx} 
  A.~A.~Korsheninnikov {\it et al.},
  Phys.\ Lett.\ B {\bf 343}, 53 (1995).




\end{thebibliography}
\end{document}